\documentclass[twocolumn,preprintnumbers,amsmath,amssymb,superscriptaddress]{revtex4-2}
\usepackage{quiver}
\usepackage{graphicx}
\usepackage{dcolumn}
\usepackage{bm}
\usepackage{amsthm}
\usepackage{amsmath}
\usepackage{amssymb}
\usepackage{tikz}
\usetikzlibrary{arrows}
\usepackage{float}

\theoremstyle{definition}

\theoremstyle{remark}

\theoremstyle{example}

\theoremstyle{conjecture}

\newcommand{\im}{{\rm Im}\,}

\newcommand {\R} {\mathbb R}
\newcommand {\C} {\mathbb C}

\newcommand {\N} {\mathbb N}

\def\bea{\begin{eqnarray}}
\def\eea{\end{eqnarray}}

\renewcommand\theequation{\arabic{equation}}
\newcommand{\bzero}{\mbox{\normalfont\bfseries 0}}

\begin{document}

\preprint{APS/123-QED}

\title{Persistent Dirac for molecular representation}

\author{JunJie Wee}
\email{weej0019@e.ntu.edu.sg}
\affiliation{%
	Division of Mathematical Sciences, School of Physical and Mathematical Sciences, Nanyang Technological University (NTU), 637371 Singapore
}%
\author{Ginestra Bianconi}
\email{ginestra.bianconi@gmail.com}
\affiliation{%
	School of Mathematical Sciences, Queen Mary University of London, London, E1 4NS, United Kingdom and The Alan Turing Institute, London, NW1 2DB, United Kingdom
}%
\author{Kelin Xia}%
\email{xiakelin@ntu.edu.sg}
\affiliation{%
	Division of Mathematical Sciences, School of Physical and Mathematical Sciences, Nanyang Technological University (NTU), 637371 Singapore
}%

\date{\today}

\begin{abstract}
Molecular representations are of fundamental importance for the modeling and analysis of molecular systems. Representation models and in general approaches based on topological data analysis (TDA) have demonstrated great success in various steps of drug design and materials discovery. Here we develop a mathematically rigorous computational framework for molecular representation based on the persistent Dirac operator. The properties of the spectrum of the discrete weighted and unweighted Dirac matrices are systemically discussed and used to demonstrate the geometric and topological properties of both non-homology and homology eigenvectors of real molecular structures. This  allows us to asses the influence of weighting schemes on the information encoded in the Dirac eigenspectrum. A series of physical persistent attributes, which characterize the spectrum of the Dirac matrices across a filtration, are proposed and used as efficient molecular fingerprints. Finally, our persistent Dirac-based model is used for clustering molecular configurations from nine types of organic-inorganic halide perovskites. We found that our model can cluster the structures very well, demonstrating the representation and featurization power of the current approach.
\end{abstract}

\maketitle


\section{\label{sec:level1}Introduction}
Molecular representation and featurization play an essential role in physical as well as  in data-driven learning models. Given the rich interplay between structure and function of molecules, an efficient characterization of structural properties is key for extracting functional information. Various quantitative structure-activity/property relationship (QSAR/QSPR) models have been developed to establish explicit linear (or nonlinear) relations between molecular structure and function \cite{puzyn2010recent,lo2018machine}. Different molecular fingerprints have been proposed for machine learning and deep learning models in the prediction of molecular functions and properties \cite{wee2021ollivier, liu2021persistent, wang2020persistent, wee2021forman,chen2021algebraic,chen2021evolutionary}. However, despite  the great progresses, the design of highly efficient descriptors is still the bottleneck for QSAR/QSPR and learning models in the analysis of molecular data from materials, chemistry and biology \cite{puzyn2010recent,lo2018machine}.

Graph models \cite{wei2022hodge,meng2019weighted,anand2020weighted,xia:2012multiscale,xia2015multiresolution,nguyen2019agl,Xia:2015multiscale,Xia:2018mvpenm, berrone2021graph,berrone2022graph} are arguably the most widely used tools for molecular representations in molecular dynamics simulation, coarse-grained models, elastic network models, QSAR/QSPR, graph neural networks, etc. In general, a molecule (or a molecular complex) is modeled as a graph with each vertex representing an atom, an amino acid, a domain, or an entire molecule, and edge representing covalent-bond, non-covalent-bond, or more general interaction. However, graphs are designed for the characterization of pairwise interactions. To capture higher-order interactions, topological representations, such as multilayer networks \cite{bianconi2018multilayer}, simplicial complexes \cite{bianconi2021higher, petri2013topological, petri2013networks}, hypergraphs \cite{barbensi2022hypergraphs, bick2021higher}, etc, should be considered. Among them, multilayer networks have been used in the characterization of higher-order dynamics \cite{torres2020simplicial,millan2020explosive,ghorbanchian2021higher, calmon2021topological} and synchronization dynamics \cite{wu2015emergent,bianconi2016network,bianconi2017emergent}. As a generalization of graphs, simplicial complexes are made not only by $0$-simplices (nodes) and $1$-simplices (edges), but also by higher-dimensional simplices, such as $2$-simplices (triangles), $3$-simplices (tetrahedron), etc. Hence, higher-order networks and simplicial complexes can describe the many-body interactions among the atoms of a molecule. Hypergraphs are a further generalization of simplicial complexes. An hypergraph is composed of hyperedges, which are formed by a set of vertices. Recently, simplicial complexes and hypergraphs have been used in molecular representations and have allowed improved performance of drug design algorithms, in particular, in the protein-ligand binding affinity prediction.

Based on topological representations, molecular descriptors or fingerprints can be generated and further used as features for learning models. Recently, topological data analysis (TDA) \cite{Edelsbrunner:2002,Zomorodian:2005} and combinatorial Hodge theory based molecular descriptors have achieved great success in various steps of drug design, including protein-ligand binding affinity prediction \cite{cang:2017topologynet,cang:2017integration,nguyen:2017rigidity,cang2018integration,nguyen2019agl,meng2021persistent, Liu2020hypergraph,liu2021persistent}, protein stability change upon mutation prediction \cite{cang:2017analysis,cang:2018representability}, toxicity prediction \cite{wu:2018quantitative}, solvation free energy prediction \cite{wang2016automatic,wang2018breaking}, partition coefficient and aqueous solubility \cite{wu2018topp}, and binding pocket detection \cite{zhao2018protein}. These models have also demonstrated great advantages over traditional molecular representations in D3R Grand challenge \cite{nguyen2018mathematical,nguyen2019mathdl}. Mathematically, the key idea of TDA is extract topological information by investigating persistent homology, which tracks the change of homology generators (i.e., Betti numbers) from simplicial complexes over a filtration process. In particular, the topological invariant  Betti numbers can be obtained from the kernel of combinatorial Hodge Laplacians (HL) matrix. Interestingly, also the Forman Ricci curvature can be obtained via the Bochner-Weitzenb\"ock decomposition of HL matrix \cite{wee2021forman}. The great success of TDA and combinatorial Hodge theory based molecular descriptors in learning models is due to their characterization of structures with intrinsic invariants, including Betti numbers and Ricci curvatures. These intrinsic descriptors are well defined mathematical observables that characterize fundamental topological and geometrical properties of real datasets, thus they have an excellent transferability for learning models.

Inspired by the success of Hodge Laplacian matrix in molecular sciences, here we propose persistent Dirac based molecular representation and fingerprint.  
The discrete Dirac operator\cite{Bianconi_2021,calmon2023dirac,post2009first,lloyd2016quantum,ameneyro2022quantumph,crane2011spin,bianconi2022dirac}
is a first-order differential operator which can be interpreted as the square root of Hodge Laplace operator. This operator has been developed on graphs and simplicial complexes and used in TDA and for investigating dynamics of topological signals \cite{giambagli2022diffusion,calmon2023dirac,calmon2023local,calmon2022dirac}. Moreover, the persistent Dirac model can be used in the quantum algorithm of persistent homology \cite{lloyd2016quantum,ameneyro2022quantumph,ameneyro2022quantum}. Here we present a rigorous mathematical theory for persistent Dirac through the commutative diagram of discrete Dirac operator over a filtration process. The commutative diagram is similar to the ones in persistent spectral graph \cite{wang2021hermes,wang2020persistent}, persistent Hodge Laplacian \cite{memoli2022persistent}, and persistent sheaf Laplacian \cite{wei2021persistent,memoli2022persistent}. Further, we develop a series of persistent attributes from persistent Dirac, and use them as descriptors to characterize molecular structures.

Our work starts with a systematically study of the spectrum of the discrete Dirac matrices. In particular, we identify the geometric and topological properties of both non-homology and homology eigenvectors for molecular structures. We generalize these results to weighted simplicial complexes on top of which the weighted Dirac operator \cite{baccini2022weighted} is carefully defined. In particular, here we analyse the influence of weighting schemes on the spectral properties of molecular structures.  The persistent Dirac is then introduced and is  employed for the clustering of molecular configurations from the molecular dynamic simulations of nine types of organic-inorganic halide perovskites (OIHP). By the comparison with several existing models, we show that our model is highly efficient in clustering the structure configurations. This demonstrates the great potential of our persistent Dirac-based fingerprints in molecular representation and featurization.

The paper is organized as follows. Sec.~\ref{sec:level2} is devoted for Hodge Laplacian model. It covers general concepts including simplicial complexes, chain groups, boundary operators, and Hodge Laplacian. In Sec.~\ref{sec:level3}, persistent Dirac model is present. The eigenspectrum information for (weighted) Dirac matrix and persistent attributes from persistent Dirac are discussed in detailed. Sec.~\ref{sec:level4} is for the application of the persistent Dirac based fingerprints. The paper ends with an conclusion.

\section{\label{sec:level2}Hodge Laplacian}

\subsection{Simplicial Complex}

Generally speaking, a simplicial complex can be viewed as a higher-dimensional generalization of graphs. A $p$-dimensional simplicial complex is formed by simplices of dimension up to $p$. Every $p$ dimensional simplex consists of a  set of $p+1$ vertices and this set can be viewed geometrically as a point ($0$-simplex), an edge ($1$-simplex), a triangle ($2$-simplex), a tetrahedron ($3$-simplex), etc.

More precisely, a $p$-simplex $\sigma^p = \{v_0, v_1, v_2, \cdots, v_p\}$ is defined as a convex hull formed by its $p+1$ affinely independent points $v_0, v_1, v_2, \cdots, v_p$:
\[
\sigma^p = \bigg\{\lambda_0v_0+\lambda_1v_1+\cdots+\lambda_pv_p \bigg|\sum_{i=0}^p \lambda_i = 0;\forall i, 0\leq \lambda_i \leq 1\bigg\}.
\]

The $i$-dimensional face of $p$-dimensional simplex $\sigma^p$ (indicated with $i<p$) is the convex hull formed by $i+1$ vertices belonging to the set of $p+1$ points $\{v_0, v_1, v_2, \cdots, v_p\}$. The simplices are basic components of a simplicial complex.

A simplicial complex $\mathcal{K}$ is a finite set of simplices that satisfy two essential conditions:
\begin{itemize}
	\item Any face of a simplex from $\mathcal{K}$ is also in $\mathcal{K}$.
	\item The intersection of any two simplices in $\mathcal{K}$ is either empty or formed by shared faces.
\end{itemize}
Here and in the following we indicate with $n_p$ the number of $p$-simplices belonging to the simplicial complex $\mathcal{K}$.
The most commonly used simplical complexes include \v{C}ech complex, Vietoris-Rips complex, Alpha complex, Cubical complex, Morse complex, etc. \cite{vaccarino2022persistent}.

Two $p$-dimensional simplices $\sigma_1$ and $\sigma_2$ in a simplicial complex $\mathcal{K}$,  are simplex neighbors if
\begin{itemize}
	\item[(i)] $\sigma_1$ and $\sigma_2$ share a $(p+1)$-simplex $\mu$, that is, there exists a $\mu$ in $\mathcal{K}$ such that $\mu > \sigma_1$ and $\mu >\sigma_2$.
	\item[(ii)] $\sigma_1$ and $\sigma_2$ share a $(p-1)$-simplex $\gamma$, that is, there exists a $\gamma$ in $\mathcal{K}$ such that $\gamma < \sigma_1$ and $\gamma < \sigma_2$.
\end{itemize}
If either condition is satisfied, as long as  both conditions do not hold at the same time,  $\sigma_1$ and $\sigma_2$ are called parallel simplex neighbors. Here $\sigma_1$ and $\sigma_2$ are called upper adjacent neighbors and denoted as $\sigma_1 \frown \sigma_2$, if they satisfy condition (i). They are lower adjacent neighbors and denoted as $\sigma_1 \smile \sigma_2$ if they satisfy condition (ii).

In addition, the $d$-skeleton of a $p$-dimensional simplicial complex is the simplicial complex consisting of simplices up to dimension $d$, where $0\leq d\leq p$. The $1$-skeleton of a simplicial complex is always the graph of simplicial complex.

\subsection{Homology}
In homology, a $p$-dimensional oriented simplex $\sigma^p$ is the set of ordered $p+1$ nodes $[v_0, v_1, \cdots, v_p]$. For example, an oriented $1$-simplex $\sigma^1 =[v_0, v_1]$ has the opposite sign of the oriented 1-simplex $[v_1, v_0]$. In other words, 
\[
[v_i, v_j] = -[v_j, v_i].
\]
Similarly, this orientation can be written for higher-order simplices in the following way,
\[
[v_0, v_1, \cdots, v_p] = (-1)^{\alpha(\pi)} [v_{\pi(0)}, v_{\pi(1)}, \cdots, v_{\pi(p)}],
\]
where $\alpha(\pi)$ refers to the parity of the permutation $\pi$. In this paper, we consider the orientation induced by node labels, i.e. for every simplex in a simplicial complex, we assign a positive orientation to the one provided by the increasing set of node labels.

For an oriented simplicial complex $\mathcal{K}$, its $p$-dimensional chain group $C_p(\mathcal{K})$ is composed by linear combination of positively oriented $p$-simplices in $\mathcal{K}$. Let $[v_0, v_1, \cdots v_p]$ indicate the generic positively oriented $p$-simplex $\sigma^p\in \mathcal{K}$. We notice that the set of simplices $\sigma_p$ constitute a basis for the $p$-dimensional chains $C_p(\mathcal{K})$. Therefore any $p$-chain $f_1\in C_p(\mathcal{K})$ can be written in a unique way as 
\begin{equation} 
f_1=\sum_{i=1}^{n_p}c_i\sigma^i.
\end{equation}

The weighted boundary operator $\overline{\partial}_p: C_p \rightarrow C_{p-1}$ can be determined by its action on  any  given $\sigma^p \in \mathcal{K}$:
\begin{equation*}
\overline{\partial}_p(\sigma^p) = a_p\sum_{i=0}^p (-1)^i [v_0, v_1, \cdots, \hat{v}_i, \cdots, v_p].
\end{equation*}
Here $a_p$ is a constant in $\mathbb{R}^+$ dependent on $p$ and the boundary of $p$-simplex is made of $(p-1)$-simplices $[v_0, v_1, \cdots, \hat{v}_i, \cdots, v_p]$, where $\hat{v_i}$ means that $v_i$ has been removed from the sequence $v_0, \cdots, v_p$. It is also well-known that $\overline{\partial}_{p-1}\overline{\partial}_{p} = 0$. The unweighted boundary operator can be obtained by setting $a_p=1$. In other words, the unweighted boundary operator $\partial_p: C_p \rightarrow C_{p-1}$ for a given $\sigma^p \in \mathcal{K}$ is defined as
\begin{equation*}
\partial_p(\sigma^p) = \sum_{i=0}^p (-1)^i [v_0, v_1, \cdots, \hat{v}_i, \cdots, v_p].
\end{equation*}

For an oriented simplicial complex $\mathcal{K}$, its two oriented $p$-dimensional simplices $\sigma_1$ and $\sigma_2$ are similarly oriented and denoted as $\sigma_1 \sim \sigma_2$, if they are lower adjacent and have the same sign on the common lower $(p-1)$-simplex. Two simplex $\sigma_1$ and $\sigma_2$ are dissimilarly oriented and denoted as $\sigma_1 \nsim \sigma_2$, if they are lower adjacent but have different signs on the common lower $(p-1)$-simplex.

The $p$-th cycle group $Z_p$ is defined as,
\begin{equation*}
Z_p =\operatorname{ker}(\overline{\partial}_{p}) = \{c \in C_p | \overline{\partial}_{p}(c) = 0\},
\end{equation*}
and $p$-th boundary group $B_p$ is,
\begin{equation*}
B_p =\operatorname{im}(\overline{\partial}_{p+1}) = \{c\in C_p | \exists d \in C_{p+1} : c = \overline{\partial}_{p+1}(d) \}.
\end{equation*}
The $p$-th homology group is defined as $H_p = Z_p/B_p$. Its rank is $p$-th Betti number that satisfies
\begin{equation*}
\beta_p = \text{rank } H_p = \text{rank } Z_p - \text{rank } B_p.
\end{equation*}

With the boundary operators, we have chain complexes

$$\cdots \xrightarrow{\overline{\partial}_{p+2}} C_{p+1}\xrightarrow{\overline{\partial}_{p+1}} C_{p}\xrightarrow{\overline{\partial}_{p}} C_{p-1}\xrightarrow{\overline{\partial}_{p-1}} \cdots $$

The adjoint of $\overline{\partial}_p$, which is
\[
\overline{\partial}_{p}^*:C_{p-1}\to C_{p},
\]
satisfies the inner product relation $\langle \overline{\partial}_p(f), g \rangle=\langle f, \overline{\partial}_{p}^*(g) \rangle,$ for every $f \in C_{p}$, $g \in C_{p-1}$. It is used in the weighted Hodge Laplacian.

\subsection{Weighted Hodge Laplacian and Hodge Decomposition}

The $p$-dimensional weighted Hodge Laplacian $\Delta_p : C_p \rightarrow C_p$ is defined as follows:
\[
\Delta_p = \begin{cases}
\overline{\partial}_1\circ \overline{\partial}_1^*, & \text{if } p=0.\\
\overline{\partial}_{p}^*\circ \overline{\partial}_{p} + \overline{\partial}_{p+1}\circ \overline{\partial}_{p+1}^*, & \text{if } p\geq 1.
\end{cases}
\]
The special case where $p = 0$ is the well-known graph Laplacian.

Computationally, the information for weighted boundary operators acting from finite dimensional chain groups $C_p$ to $C_{p-1}$ can be stored efficiently in matrix representations. As matrix representations, the weighted boundary operators and its adjoint satisfies $\overline{\partial}_{p}^\top= \overline{\partial}_{p}^*$.

More specifically, let $n_{p-1}$ and $n_p$ be the number of $(p-1)$-simplices and $p$-simplices respectively in a simplicial complex $\mathcal{K}$.
The $n_{p-1}\times n_p$ weighted boundary matrix $\overline{\mathbf{B}}_p$ has entries defined as follows:
\begin{equation*}\label{eqn:boundary}
\overline{\mathbf{B}}_p (i,j) = \left\{ \begin{array}{ll}
a_p, & \text{if } \sigma_i^{p-1} < \sigma_j^p, \sigma_i^{p-1} \sim \sigma_j^p.\\
-a_p, & \text{if } \sigma_i^{p-1} < \sigma_j^p, \sigma_i^{p-1} \nsim \sigma_j^p.\\
0, & \text{if } \sigma_i^{p-1} \nless \sigma_j^p.
\end{array} \right.
\end{equation*}
where $1\leq i\leq n_{p-1}$ and $1\leq j\leq n_p$. Here, $\sigma_i^{p-1} < \sigma_j^p$ represents the $i$-th $(p-1)$-simplex $\sigma_i^{p-1}$ is a face of $j$-th $p$-simplex  $\sigma_j^p$ and $\sigma_i^{p-1} \sim \sigma_j^p$ indicates the coefficient of $\sigma_i^{p-1}$ in $\overline{\partial}_{p}(\sigma_j^p)$ is $a_p$. Likewise, $\sigma_i^{p-1} \nless \sigma_j^p$ means that $\sigma_i^{p-1}$ is not a face of $\sigma_j^p$ and $\sigma_i^{p-1} \nsim \sigma_j^p$ indicates that the coefficient of $\sigma_i^{p-1}$ in $\overline{\partial}_{p}(\sigma_j^p)$ is $-a_p$.

Since the unweighted boundary operator $\partial_{p}=\frac{1}{a_p}\overline{\partial}_p$, note that an unweighted boundary matrix can be similarly written as
\begin{equation}\label{eqn:boundary_matrix}
	\mathbf{B}_p =\frac{1}{a_p}\overline{\mathbf{B}}_p.
\end{equation}

Using the weighted boundary matrices, the lower and upper weighted Hodge Laplacians can be defined as
$\overline{\mathbf{L}}_p^{\text{down}} = \overline{\mathbf{B}}_p^\top\overline{\mathbf{B}}_p$ and $\overline{\mathbf{L}}_p^{\text{up}} = \overline{\mathbf{B}}_{p+1}\overline{\mathbf{B}}_{p+1}^\top$ respectively. More specifically, the entries of $\overline{\mathbf{L}}_p^{\text{down}}$  ($p>0$) are as follows,
\begin{equation*}
\overline{\mathbf{L}}_p^{\text{down}} (i,j) = \left\{
\renewcommand{\arraystretch}{1.5}
\begin{array}{ll}
a_p^2(p+1), & i=j. \\
a_p^2, & i\ne j, \sigma^p_i \smile \sigma^p_j, \sigma^p_i \sim \sigma^p_j. \\
-a_p^2, & i\ne j, \sigma^p_i \smile \sigma^p_j, \sigma^p_i \nsim \sigma^p_j. \\
0, & i\ne j \text{ and } \sigma^p_i \not\smile \sigma^p_j.
\end{array} \right.
\end{equation*}
For $p>0$ the matrix elements of the Hodge Laplacian $\overline{\mathbf{L}}_p^{\text{up}}$  are given by 
\begin{equation*}
\overline{\mathbf{L}}_p^{\text{up}} (i,j) = \left\{
\renewcommand{\arraystretch}{1.5}
\begin{array}{ll}
a_{p+1}^2 d(\sigma^p_i), & i=j. \\
-a_{p+1}^2, & i\ne j, \sigma^p_i \frown \sigma^p_j, \sigma^p_i \sim \sigma^p_j. \\
a_{p+1}^2, & i\ne j, \sigma^p_i \frown \sigma^p_j, \sigma^p_i \nsim \sigma^p_j. \\
0, & i\ne j \text{ and } \sigma^p_i \not\frown \sigma^p_j.
\end{array} \right.
\end{equation*}
Here $d(\sigma^p_i)$ denotes the number of cofaces with dimension $p+1$ of simplex $\sigma^p_i$. Note that all the entries of $\overline{\mathbf{L}}_0^{\text{down}}$ are zero since $0$-simplices have no lower adjacent neighbors. Further, $\sigma^p_i \smile \sigma^p_j$ refers to $\sigma^p_i$ and $\sigma^p_j$ being lower adjacent neighbors while $\sigma^p_i \frown \sigma^p_j$ refers to $\sigma^p_i$ and $\sigma^p_j$ being upper adjacent neighbors.

These matrices, i.e., $\overline{\mathbf{B}}_p$, $\overline{\mathbf{L}}_p^{\text{up}}$ and $\overline{\mathbf{L}}_p^{\text{down}}$, have various interesting properties as follows (see \cite{horak2013spectra} or Appendix A for proofs).
		\begin{itemize}
			\item [(i)] $\ker \overline{\mathbf{B}}_p = \ker \overline{\mathbf{L}}_p^{\text{down}}$.
			\item [(ii)] $\ker \overline{\mathbf{B}}_{p}^\top = \ker \overline{\mathbf{L}}_{p-1}^{\text{up}}$.
			\item [(iii)] $\lambda$ is a non-zero eigenvalue of $\overline{\mathbf{L}}_{p}^{\text{down}}$ with corresponding eigenvector $v$ if and only if $\lambda$ is a non-zero eigenvalue of $\overline{\mathbf{L}}_{p-1}^{\text{up}}$ with corresponding eigenvector $\overline{\mathbf{B}}_pv$. Hence, $\overline{\mathbf{L}}_{p-1}^{\text{up}}$ and $\overline{\mathbf{L}}_{p}^{\text{down}}$ always have the same non-zero eigenvalues.
			\item [(iv)] $v \in \ker \overline{\mathbf{L}}_{p}^{\text{down}}$ if and only if $\overline{\mathbf{B}}_pv \in \ker \overline{\mathbf{L}}_{p-1}^{\text{up}}$.
			\item [(v)] $\operatorname{im} \overline{\mathbf{L}}_p^{\text{up}} \subset \ker \overline{\mathbf{L}}_p^{\text{down}}$.
			\item [(vi)] $\operatorname{im} \overline{\mathbf{L}}_p^{\text{down}} \subset \ker \overline{\mathbf{L}}_p^{\text{up}}$.
			\item [(v)] { $\ker \overline{\mathbf{B}}_p^\top = (\operatorname{im } \overline{\mathbf{B}}_p)^\perp$ }
		\end{itemize}

The $p^\text{th}$ weighted combinatorial Laplacian $\overline{\mathbf{L}}_p$ is defined as $\overline{\mathbf{L}}_p=\overline{\mathbf{B}}_p^\top\overline{\mathbf{B}}_p + \overline{\mathbf{B}}_{p+1}\overline{\mathbf{B}}_{p+1}^\top $. Note that $\overline{\mathbf{L}}_0=\overline{\mathbf{B}}_1\overline{\mathbf{B}}_1^\top$. The matrix elements of the Hodge Laplacians  $\overline{\mathbf{L}}_p$ with $p=0$ are given by 
\begin{equation*}
\overline{\mathbf{L}}_0 (i,j) = \left\{ \begin{array}{ll}
a_1^2d(\sigma^0_i), & i=j. \\
-a_1^2, & i\ne j, \sigma^0_i \frown \sigma^0_j.\\
0, & i\ne j, \sigma^0_i \not\frown \sigma^0_j.
\end{array} \right.
\end{equation*}
while the matrix elements for $p>0$ can be expressed as 
\begin{widetext}
\begin{equation*}\small
\overline{\mathbf{L}}_p (i,j)=\left\{ \begin{array}{ll}
a_{p+1}^2d(\sigma^p_i)+a_p^2(p+1), & i=j. \\
a_p^2 - a_{p+1}^2, & i\ne j, \sigma^p_i \frown \sigma^p_j, \sigma^p_i \smile \sigma^p_j, \sigma^p_i \sim \sigma^p_j. \\
a_{p+1}^2 - a_p^2, & i\ne j, \sigma^p_i \frown \sigma^p_j, \sigma^p_i \smile \sigma^p_j, \sigma^p_i \not\sim \sigma^p_j. \\
a_{p}^2, & i\ne j, \sigma^p_i \not\frown \sigma^p_j, \sigma^p_i \smile \sigma^p_j, \sigma^p_i \sim \sigma^p_j. \\
-a_{p}^2, & i\ne j, \sigma^p_i \not\frown \sigma^p_j, \sigma^p_i \smile \sigma^p_j, \sigma^p_i \nsim \sigma^p_j. \\
0, & i\ne j \text{ and }\sigma^p_i \not\smile \sigma^p_j.
\end{array} \right.
\end{equation*}
\end{widetext}
It follows from Eq. \eqref{eqn:boundary_matrix} that the lower and upper unweighted Hodge Laplacians can be written as $\mathbf{L}_p^{\text{down}} = \mathbf{B}_p^\top\mathbf{B}_p$ and $\mathbf{L}_p^{\text{up}} = \mathbf{B}_{p+1}\mathbf{B}_{p+1}^\top$ respectively. Hence, the $p^\text{th}$ unweighted combinatorial Laplacian $\mathbf{L}_p = \mathbf{L}_p^{\text{down}} + \mathbf{L}_p^{\text{up}}$ have elements given by 
\begin{equation*}
\mathbf{L}_0 (i,j) = \left\{ \begin{array}{ll}
d(\sigma^0_i), & i=j. \\
-1, & i\ne j, \sigma^0_i \frown \sigma^0_j.\\
0, & i\ne j, \sigma^0_i \not\frown \sigma^0_j.
\end{array} \right.
\end{equation*}
for $p=0$ while for $p>0$ the matrix elements of the Hodge Laplacian are given by 
\begin{widetext}
\begin{equation*}\small
\mathbf{L}_p (i,j)=\left\{ \begin{array}{ll}
d(\sigma^p_i)+p+1, & i=j. \\
1, & i\ne j, \sigma^p_i \not\frown \sigma^p_j, \sigma^p_i \smile \sigma^p_j, \sigma^p_i \sim \sigma^p_j. \\
-1, & i\ne j, \sigma^p_i \not\frown \sigma^p_j, \sigma^p_i \smile \sigma^p_j, \sigma^p_i \nsim \sigma^p_j. \\
0, & i\ne j \text{ and either } \sigma^p_i \frown \sigma^p_j \text{ or } \sigma^p_i \not\smile \sigma^p_j.
\end{array} \right.
\end{equation*}
\end{widetext}

It is well-known that $\lambda$ is a non-zero eigenvalue of $\overline{\mathbf{L}}_p$ if and only if $\lambda$ is an non-zero eigenvalue of $\overline{\mathbf{L}}_p^{\text{down}}$ or $\overline{\mathbf{L}}_p^{\text{up}}$. The multiplicity of the zero eigenvalues of $\overline{\mathbf{L}}_p$ corresponds to the $p$th Betti number as follows,
\begin{equation*}\label{eqn:kernel_lp}
	\dim \ker \overline{\mathbf{L}}_p = \beta_p = \dim \ker \overline{\mathbf{L}}_p^{\text{down}} - \dim \operatorname{im} \overline{\mathbf{L}}_p^{\text{up}}
\end{equation*}
where $\beta_p$ is also the $\text{rank }H_p$ (see \cite{horak2013spectra} or Appendix B).

Further, $\dim \ker \overline{\mathbf{L}}_p^{\text{down}}$ can be written as:
\begin{align}\label{eq:kernel_lp_down}
\dim \ker \overline{\mathbf{L}}_p^{\text{down}} &= \beta_p + \dim \operatorname{im} \overline{\mathbf{L}}_p^{\text{up}}\nonumber \\
&= \beta_p + \dim C_p - \dim \ker \overline{\mathbf{L}}_p^{\text{up}}\nonumber \\
&= \beta_p + \dim C_p - \dim \ker \overline{\mathbf{B}}_{p+1}^\top\nonumber \\
&= \beta_p + \text{rank } \overline{\mathbf{B}}_{p+1}^\top.
\end{align}
This means that the number of zero eigenvalues in $\overline{\mathbf{L}}_p^{\text{down}}$ is equal to the sum of $\text{rank } \overline{\mathbf{B}}_{p+1}^\top$ and $\beta_p$. { Furthermore, since $H_p = Z_p/B_p= \ker \overline{\mathbf{B}}_p/\text{im }\overline{\mathbf{B}}_{p+1}$, then
\[
\ker \overline{\mathbf{L}}_p \cong H_p = \ker \overline{\mathbf{B}}_p \cap (\text{im } \overline{\mathbf{B}}_{p+1})^{\perp}.
\]
Here, $(\text{im } \overline{\mathbf{B}}_{p+1})^{\perp}$ is the orthogonal complement of $\text{im } \overline{\mathbf{B}}_{p+1}$. In fact, $(\text{im } \overline{\mathbf{B}}_{p+1})^{\perp} = \ker \overline{\mathbf{B}}_{p+1}^\top$. This is true because from (v), $\ker \overline{\mathbf{B}}_{p}^\top = (\text{im } \overline{\mathbf{B}}_{p})^\perp$ (see Appendix A for the proof). Hence, this gives
\[
\ker \overline{\mathbf{B}}_p \cap (\text{im } \overline{\mathbf{B}}_{p+1})^{\perp} = \ker \overline{\mathbf{B}}_p \cap \ker \overline{\mathbf{B}}_{p+1}^\top.
\]
In other words, we have
\[
\ker \overline{\mathbf{L}}_p = \ker \overline{\mathbf{B}}_p \cap \ker \overline{\mathbf{B}}_{p+1}^\top.
\]
This means that for every $v\in \ker \overline{\mathbf{L}}_p$,
\[
v\in \ker \overline{\mathbf{B}}_p \cap \ker \overline{\mathbf{B}}_{p+1}^\top.
\]
}

This is exactly the Hodge decomposition which states that a $p$-th chain group $C_p$ of a simplicial complex $\mathcal{K}$ admits the following orthogonal direct sum decomposition:
	\begin{equation*}
	C_p = \rlap{$\underbrace{\phantom{\operatorname{im}(\overline{\mathbf{B}}_{p+1})}}_{\operatorname{im}\overline{\mathbf{L}}_p^{\text{down}}}$} \rlap{$\overbrace{ \phantom{\operatorname{im}(\overline{\mathbf{B}}_{p+1})\oplus\ker(\overline{\mathbf{L}}_p)}}^{\ker \overline{\mathbf{L}}_p^{\text{up}}=\ker \overline{\mathbf{B}}_{p+1}^\top}$} \operatorname{im}(\overline{\mathbf{B}}_{p+1})\oplus \underbrace{\ker(\overline{\mathbf{L}}_p) \oplus\operatorname{im}(\overline{\mathbf{B}}_{p}^\top)}_{\ker\overline{\mathbf{L}}_p^{\text{down}}=\ker \overline{\mathbf{B}}_p}\llap{$\overbrace{\phantom{\operatorname{im}(\overline{\mathbf{B}}_{p}^\top)}}^{\operatorname{im}\overline{\mathbf{L}}_p^{\text{up}}}$},
	\end{equation*}
	where $\ker(\overline{\mathbf{L}}_p) = \ker \overline{\mathbf{B}}_p \cap \ker \overline{\mathbf{B}}_{p+1}^\top$.

The vector space of edge flows $C_1$ admits the following orthogonal sum decomposition in Helmholtz-Hodge Decomposition:
\[
C_1=\operatorname{im}(\text{curl}^*)\oplus\ker(\overline{\mathbf{L}}_1)\oplus\operatorname{im}(\text{grad}),
\]
where
\[
\ker(\overline{\mathbf{L}}_1) = \ker(\text{curl}) \cap \ker(\text{div}).
\]
It is worth mentioning that such flows have also been extended to five component decompositions with edge and face vector fields \cite{zhao20193d}, applied to the protein B-factor prediction problems via Hodge theory \cite{chen2021evolutionary} and also in de Rham-Hodge biomolecular data analysis \cite{zhao2020rham, wei2022hodge}.

\section{\label{sec:level3}Persistent Dirac}

\subsection{Discrete Dirac models}

\paragraph{Weighted Dirac matrix}

Recently, weighted Dirac matrices have been proposed based on a weighted simplicial complex \cite{baccini2022weighted}. For a $d$-dimensional weighted simplicial complex $\mathcal{K}$, let us define the $n_{p} \times n_{p}$ metric matrix $\mathbf{G}_{p}$ ($0\leq p \leq d$) to be a diagonal matrix with positive entries. For any two $p$-chains $f_1=\sum_{i=1}^{n_p}c_i\sigma^i$ and $f_2=\sum_{i=0}^{n_p}d_i\sigma^i$ in $C_p$, the matrix $\mathbf{G}_{p}$ can be used to define the weighted inner product
	\begin{equation}\label{eqn:weighted_innerprod}
	\left\langle f_1, f_2\right\rangle = \sum_{i=1}^{n_{p}} \mathbf{G}_{p}(\sigma^{i},\sigma^{i})c_{i}d_i = (\mathbf{f}_1)^\top\mathbf{G}_{p}(\mathbf{f}_2),
	\end{equation}
	where $(\mathbf{f}_1)^\top=[c_1, c_2, \cdots, c_p]$ and $(\mathbf{f}_2)^\top=[d_1, d_2, \cdots, d_p]$.
	
	Recall from Eq. \eqref{eqn:boundary_matrix} that the weighted boundary operator can be represented by a matrix $\overline{\mathbf{B}}_p =a_p\mathbf{B}_p$ where $\mathbf{B}_p$ is the unweighted boundary matrix. If $a_p=1$, then $\overline{\mathbf{B}}_p$ reduces to the adjoint operator of $\mathbf{B}_p^\top$.
	Formally, for any $p$-chain $f$ and any $(p-1)$-chain $g$, the adjoint operator $\overline{\mathbf{B}}_p^*$
	satisfies
	\begin{equation}\label{eqn:inner-prod}
	\langle f, \overline{\mathbf{B}}_p^*g \rangle = \langle \overline{\mathbf{B}}_pf, g\rangle.
	\end{equation}
	From the inner product relation \eqref{eqn:inner-prod}, an explicit expression of $\overline{\mathbf{B}}_p^*$ can be deduced in terms of $\overline{\mathbf{B}}_p$ and the matrices $\mathbf{G}_p$. Based on the weighted inner product definition \eqref{eqn:weighted_innerprod}, this gives
	\begin{equation*}
	(\mathbf{f})^\top\mathbf{G}_p\overline{\mathbf{B}}_p^*(\mathbf{g}) = (\mathbf{f})^\top\overline{\mathbf{B}}_p^\top\mathbf{G}_{p-1}(\mathbf{g}).
	\end{equation*}
	Since the expression is true for any arbitrary $\mathbf{f}$ and $\mathbf{g}$, this implies
	\begin{equation*}
	\mathbf{G}_p\overline{\mathbf{B}}_p^* = \overline{\mathbf{B}}_p^\top\mathbf{G}_{p-1}.
	\end{equation*}
	Hence, the following becomes an explicit expression for the adjoint operator $\overline{\mathbf{B}}_p^*$:
	\begin{equation}\label{eqn:adjoint}
	\overline{\mathbf{B}}_p^* = \mathbf{G}_{p}^{-1} \overline{\mathbf{B}}_p^\top \mathbf{G}_{p-1}.
	\end{equation}
	Here $\overline{\mathbf{B}}_p^*$ \cite{horak2013spectra} is the adjoint of the weighted boundary operator \cite{wu2018weighted,meng2019weighted}. It is important to note that if the metric matrices $\mathbf{G}_p$ are the identity matrices, the above expression then reduces to the transpose of the boundary operator multiplied by the constant $a_p$,
	\begin{equation}\label{eqn:adjoint2}
	 \overline{\mathbf{B}}_p^* = \overline{\mathbf{B}}_p^\top = a_p\mathbf{B}_p^\top.
	\end{equation}
	This also means the transpose of adjoint operator, i.e. $(\overline{\mathbf{B}}_p^*)^\top$, is equal to $\overline{\mathbf{B}}_p$ only if $\mathbf{G}_p$ are identity matrices.
	To see this, apply the transpose to both sides of Eq. \eqref{eqn:adjoint} and obtain the expression
	\begin{equation}\label{eqn:transpose_adjoint}
	(\overline{\mathbf{B}}_p^*)^\top = \mathbf{G}_{p-1}^{-1}\overline{\mathbf{B}}_p\mathbf{G}_{p}=a_p\mathbf{G}_{p-1}^{-1}\mathbf{B}_p\mathbf{G}_{p}.
	\end{equation}
	The matrices $(\overline{\mathbf{B}}_p^*)^\top$ and $\overline{\mathbf{B}}_p^*$ can then be used to construct the following weighted Dirac matrix \eqref{eqn:dirac_weighted}.

	For a simplicial complex $\mathcal{K}$ with {$n_{p} \times n_{p-1}$} adjoint operators $\overline{\mathbf{B}}_p^*$ where $n_{p-1}$ is the number of $(p-1)$-simplices and $n_p$ is the number of $p$-simplices in $\mathcal{K}$, the weighted Dirac matrix $\overline{\mathbf{D}}_p$ is
	\begin{widetext}
		\begin{equation}
		\label{eqn:dirac_weighted}
		\renewcommand{\arraystretch}{1.5}
		\overline{\mathbf{D}}_p=
		\begin{bmatrix}
		\bzero_{n_0\times n_0}    & (\overline{\mathbf{B}}_1^*)^\top    & \bzero_{n_0\times n_2}     & \cdots      & \bzero_{n_0\times n_{p}} & \bzero_{n_0\times n_{p+1}}  \\
		
		\overline{\mathbf{B}}_1^*  & \bzero_{n_1\times n_1}      & (\overline{\mathbf{B}}_2^*)^\top    & \cdots      & \bzero_{n_1\times n_{p}} & \bzero_{n_1\times n_{p+1}}  \\
		
		\bzero_{n_2\times n_0}      & \overline{\mathbf{B}}_2^*  & \bzero_{n_2\times n_2}       & \cdots    & \bzero_{n_2\times n_{p}} & \bzero_{n_2\times n_{p+1}}   \\
		
		\vdots & \vdots & \vdots & \vdots & \vdots & \vdots  \\
		
		\bzero_{n_{p}\times n_0}       & \bzero_{n_{p}\times n_1}       & \bzero_{n_{p}\times n_2}       & \cdots       &  \bzero_{n_{p} \times n_{p}}       & (\overline{\mathbf{B}}_{p+1}^*)^\top\\
		
		\bzero_{n_{p+1}\times n_0}       & \bzero_{n_{p+1}\times n_1}       & \bzero_{n_{p+1}\times n_2}       & \cdots       & \overline{\mathbf{B}}_{p+1}^* & \bzero_{n_{p+1}\times n_{p+1}}
		\end{bmatrix}.
		\end{equation}
	\end{widetext}

	In particular, we set $a_p=(p+1)^{-1/2}$ for all $p$ up to the order of the simplicial complex and consider the matrices $\overline{\mathbf{B}}_p^*$ and $(\overline{\mathbf{B}}_p^*)^\top$ in Eq. \eqref{eqn:adjoint2} and \eqref{eqn:transpose_adjoint} respectively. For $p=2$, the weighted Dirac matrix from \eqref{eqn:dirac_weighted} becomes
	\begin{widetext}
\begin{equation*}
\label{eqn:metricdirac}
\renewcommand{\arraystretch}{1.5}
{\bf D}_1=\begin{bmatrix}
\bzero_{n_0\times n_0}    & \mathbf{G}_{0}^{-1}\mathbf{B}_1\mathbf{G}_{1}/\sqrt{2}    & \bzero_{n_0\times n_2}     & \bzero_{n_0\times n_{3}}\\

\mathbf{B}_1^\top/\sqrt{2}  & \bzero_{n_1\times n_1}      & \mathbf{G}_{1}^{-1}\mathbf{B}_2\mathbf{G}_{2}/\sqrt{3}    & \bzero_{n_1\times n_{3}}\\

\bzero_{n_2\times n_0}      & \mathbf{B}_2^\top/\sqrt{3}  & \bzero_{n_2\times n_2}       & \mathbf{G}_{2}^{-1}\mathbf{B}_3\mathbf{G}_{3}/2\\

\bzero_{n_3\times n_0}      & \bzero_{n_3\times n_1}   & \mathbf{B}_3^\top/2        & \bzero_{n_3\times n_3}
\end{bmatrix}
\end{equation*}
\end{widetext}
This definition can be extended easily to higher dimensions. Note that $\mathbf{B}_p^\top/\sqrt{p+1}$ is the adjoint operator $\overline{\mathbf{B}}_p^*$ and $\mathbf{G}_{p-1}^{-1}\mathbf{B}_p\mathbf{G}_{p}/\sqrt{p+1}$ is equal to the transpose of $\overline{\mathbf{B}}_p^*$. Note that this definition of weighted Dirac is self-adjoint and with eigenvalues smaller than or equal to one. The square of the weighted Dirac also forms a diagonal block of metric Hodge Laplacian matrices
\begin{eqnarray*}
	\label{eqn:d^2} \small
	\renewcommand{\arraystretch}{1.5} {\bf D}_2^2=\begin{bmatrix}
		\mathbf{L}_{[0]}    & \bzero_{n_0\times n_1}    & \bzero_{n_0\times n_2}     & \bzero_{n_0\times n_{3}} \\
		
		\bzero_{n_1\times n_0}  & \mathbf{L}_{[1]}      & \bzero_{n_1\times n_{2}}    & \bzero_{n_1\times n_{3}} \\
		
		\bzero_{n_2\times n_0}      & \bzero_{n_2\times n_1}  & \mathbf{L}_{[2]}   & \bzero_{n_2\times n_{3}} \\
		
		\bzero_{n_{3}\times n_0}       & \bzero_{n_{3}\times n_1}       & \bzero_{n_{3}\times n_2}       & {\mathbf{L}}_{3}^{\text{down}}\\
	\end{bmatrix}
\end{eqnarray*}
where the metric Hodge Laplacian matrices are defined as
\begin{equation*}
\mathbf{L}_{[p]} = \mathbf{L}_{[p]}^{\text{down}} + \mathbf{L}_{[p]}^{\text{up}},
\end{equation*}
with
\begin{equation*}
\begin{split}
\mathbf{L}_{[p]}^{\text{down}} &= \mathbf{B}_{p}^\top\mathbf{G}_{p-1}^{-1}\mathbf{B}_{p}\mathbf{G}_{p}/(p+1),\\
\mathbf{L}_{[p]}^{\text{up}} &=
\mathbf{G}_{p}^{-1}\mathbf{B}_{p+1}\mathbf{G}_{p+1}\mathbf{B}_{p+1}^\top/(p+2).
\end{split}
\end{equation*}

Depending on the matrices $\mathbf{G}_{p}$, the weighted Dirac matrix may not always be symmetric, despite its eigenspectrum can be shown to be always real (see Appendix H).
 
For the rest of the paper, the metric matrices $\mathbf{G}_{p}$ shall be defined with each metric value for a simplex $\sigma^{p}$ to be dependent on its $(p+1)$-dimensional cofaces in the following way \cite{baccini2022weighted}:
\begin{equation*} \label{eq:weighted_matrix} \small
\mathbf{G}_{p}(\sigma^{p},\sigma^{p}) = \left\{
\begin{array}{ll}
w_{\sigma^{d}}, & p=d \\
w_{\sigma^{p}} + \displaystyle\sum_{\sigma^{p} <\sigma^{p+1}} \mathbf{G}_{p+1}(\sigma^{p+1}, \sigma^{p+1}), & 0\leq p<d.
\end{array} \right.
\end{equation*}
Here, {$w_{\sigma^{p}}>0$ is a positive weight} on $p$-simplex $\sigma^{p}$, which can be related to physical, chemical and biological properties.

\paragraph{Discrete Dirac matrix}

With the weighted Dirac matrix $\overline{\mathbf{D}}_p$, a discrete Dirac matrix is simply the special case of $\overline{\mathbf{D}}_p$ when $\mathbf{G}_p$ are identity matrices and $a_p=1$ for all $p\geq 1$.

Previously, a general Dirac matrix has been defined as \cite{Bianconi_2021, knill2013dirac, knill2013mckean}
\begin{equation}\nonumber
\mathbf{D}_p(z) =
\renewcommand{\arraystretch}{1.5}
\begin{bmatrix}
\bzero_{n_p\times n_{p}} & z\mathbf{B}_{p+1} \\
\overline{z}\mathbf{B}_{p+1}^\top & \bzero_{n_{p+1}\times n_{p+1}}
\end{bmatrix},
\end{equation}
where $z\in\C$ such that $|z|=1$. Since $|z|=1$, the typical values of $z$ occurs when $z=\overline{z}=1$ or $z=-\overline{z}=i$. In general, the parameter $z\in \C$ extends the real eigenvectors of $\mathbf{D}_p(z)$ to $\C$ while the eigenvalue remains unchanged. By taking the square of the Dirac operator, we have
\[
\mathbf{D}_p^2 (z)=
\begin{bmatrix}
\mathbf{L}_{p}^{\text{up}} &\bzero_{n_p\times n_{p+1}} &  \\
\bzero_{n_{p+1}\times n_{p}} & \mathbf{L}_{p+1}^{\text{down}}
\end{bmatrix},
\]
which implies that the eigenvalues of diagonal block real-valued Hodge-Laplacian matrices will also be the eigenvalues of $\mathbf{D}_p^2 (z)$. Since the Hodge-Laplacians are positive semi-definite symmetric matrices, the eigenvalues of $\mathbf{D}_p^2 (z)$ are non-negative as well. However, eigenvectors from the Dirac matrix may contain complex numbers.

For a simplicial complex $\mathcal{K}$ with $n_{p-1} \times n_p$ boundary matrices $\mathbf{B}_p$ where $n_{p-1}$ is the number of $(p-1)$-simplices and $n_p$ is the number of $p$-simplices in $\mathcal{K}$, the discrete Dirac matrix $\mathbf{D}_p$ \cite{knill2013mckean} is

\begin{widetext}
\begin{equation}
\label{eqn:dirac}
\renewcommand{\arraystretch}{1.5}
\mathbf{D}_p=
\begin{bmatrix}
\bzero_{n_0\times n_0}    & \mathbf{B}_1    & \bzero_{n_0\times n_2}     & \cdots      & \bzero_{n_0\times n_{p}} & \bzero_{n_0\times n_{p+1}}  \\

\mathbf{B}_1^\top  & \bzero_{n_1\times n_1}      & \mathbf{B}_2    & \cdots      & \bzero_{n_1\times n_{p}} & \bzero_{n_1\times n_{p+1}}  \\

\bzero_{n_2\times n_0}      & \mathbf{B}_2^\top  & \bzero_{n_2\times n_2}       & \cdots    & \bzero_{n_2\times n_{p}} & \bzero_{n_2\times n_{p+1}}   \\

\vdots & \vdots & \vdots & \vdots & \vdots & \vdots  \\

\bzero_{n_{p}\times n_0}       & \bzero_{n_{p}\times n_1}       & \bzero_{n_{p}\times n_2}       & \cdots       &  \bzero_{n_{p} \times n_{p}}       & \mathbf{B}_{p+1}\\

\bzero_{n_{p+1}\times n_0}       & \bzero_{n_{p+1}\times n_1}       & \bzero_{n_{p+1}\times n_2}       & \cdots       & \mathbf{B}_{p+1}^\top  & \bzero_{n_{p+1}\times n_{p+1}}
\end{bmatrix}.
\end{equation}
\end{widetext}

It is of size $\sum_{i=0}^{p+1} n_i \times \sum_{i=0}^{p+1} n_i$. Fig.~\ref{fig:dirac_example} shows a simple construction of discrete Dirac matrices \eqref{eqn:dirac} for a triangle and a tetrahedron. In Fig.~\ref{fig:dirac_example}(a), the triangle is a 2-simplex and hence the largest Dirac operator is $\mathbf{D}_1$. On the other hand, the tetrahedron in Fig.~\ref{fig:dirac_example}(b) is a 3-simplex and thus the largest Dirac operator is $\mathbf{D}_2$.

Note that by taking the square of $\mathbf{D}_p$, one would obtain a matrix with diagonal blocks of unweighted combinatorial Hodge Laplacians as shown below.
\begin{eqnarray*}
	\label{eqn:d^2} \small
	\renewcommand{\arraystretch}{1.2} {\mathbf{D}}_p^2=\begin{bmatrix}
		\mathbf{L}_0    & \bzero_{n_0\times n_1}    & \bzero_{n_0\times n_2}     & \cdots      & \bzero_{n_0\times n_{p}} & \bzero_{n_0\times n_{p+1}}  \\
		
		\bzero_{n_1\times n_0}  & \mathbf{L}_1      & \bzero_{n_1\times n_{2}}    & \cdots      & \bzero_{n_1\times n_{p}} & \bzero_{n_1\times n_{p+1}}  \\
		
		\bzero_{n_2\times n_0}      & \bzero_{n_2\times n_1}  & \mathbf{L}_2       & \cdots    & \bzero_{n_2\times n_{p}} & \bzero_{n_2\times n_{p+1}}   \\
		
		\vdots & \vdots & \vdots & \vdots & \vdots & \vdots  \\
		
		\bzero_{n_{p}\times n_0}       & \bzero_{n_{p}\times n_1}       & \bzero_{n_{p}\times n_2}       & \cdots       & \mathbf{L}_{p}      & \bzero_{n_{p}\times n_{p+1}}\\
		
		\bzero_{n_{p+1}\times n_0}       & \bzero_{n_{p+1}\times n_1}       & \bzero_{n_{p+1}\times n_2}       & \cdots       & \bzero_{n_{p+1}\times n_{p}}  & \mathbf{L}_{p+1}^{\text{down}}
	\end{bmatrix},
\end{eqnarray*}
where the unweighted  Hodge Laplacian $\mathbf{L}_{p}$,  is given by $\mathbf{L}_{p}=\mathbf{L}_{p}^{\text{down}}+ \mathbf{L}_{p}^{\text{up}}$ with $\mathbf{L}_{p}^{\text{down}}=\mathbf{B}_{p}^\top\mathbf{B}_{p}$ and $\mathbf{L}_{p}^{\text{up}}=\mathbf{B}_{p+1}\mathbf{B}_{p+1}^\top$. In our case, the last term contains only $\mathbf{L}_{p+1}^{\text{down}}$.

Recall that $\mathbf{B}_{p+1}^\top\mathbf{B}_{p+1}$ is also known as the lower Hodge Laplacian $\mathbf{L}_{p+1}^{\text{down}}$ while $\mathbf{B}_{p+2}\mathbf{B}_{p+2}^\top$ is known as the upper Hodge Laplacian $\mathbf{L}_{p+1}^{\text{up}}$.
\begin{equation*}\label{eqn:hl}
\mathbf{L}_{p+1} = \mathbf{L}_{p+1}^{\text{up}} + \mathbf{L}_{p+1}^{\text{down}}.
\end{equation*}

\subsection{\label{subsec:level2} Spectrum of the discrete Dirac operator}

\paragraph{Spectral of Dirac matrix}
Let $\mathbf{Q}_p$ be the block diagonal matrix
\begin{widetext}
\begin{eqnarray*}
	\label{eqn:I^2} \tiny
	\renewcommand{\arraystretch}{1.5}\mathbf{Q}_p=\begin{bmatrix}
		\mathbf{I}_{n_0}    & \bzero_{n_0\times n_1}    & \bzero_{n_0\times n_2}     & \cdots      & \bzero_{n_0\times n_{p}} & \bzero_{n_0\times n_{p+1}}  \\
		
		\bzero_{n_1\times n_0}  & -\mathbf{I}_{n_1}    & \bzero_{n_1\times n_{2}}    & \cdots      & \bzero_{n_1\times n_{p}} & \bzero_{n_1\times n_{p+1}}  \\
		
		\bzero_{n_2\times n_0}      & \bzero_{n_2\times n_1}  & \mathbf{I}_{n_2}       & \cdots    & \bzero_{n_2\times n_{p}} & \bzero_{n_2\times n_{p+1}}   \\
		
		\vdots & \vdots & \vdots & \vdots & \vdots & \vdots  \\
		
		\bzero_{n_{p}\times n_0}       & \bzero_{n_{p}\times n_1}       & \bzero_{n_{p}\times n_2}       & \cdots       & (-1)^{p}\mathbf{I}_{n_p}      & \bzero_{n_{p}\times n_{p+1}}\\
		
		\bzero_{n_{p+1}\times n_0}       & \bzero_{n_{p+1}\times n_1}       & \bzero_{n_{p+1}\times n_2}       & \cdots       & \bzero_{n_{p+1}\times n_{p}}  & (-1)^{p+1}\mathbf{I}_{n_{p+1}}
	\end{bmatrix},
\end{eqnarray*}
\end{widetext}
where $\mathbf{I}_{n_p}$ denotes an $n_p \times n_p$ identity matrix and $\mathbf{Q}_p$ satisfies
\begin{equation*}
\mathbf{Q}_p^2 = \mathbf{I}_{\sum_{i=0}^{p+1} n_i}.
\end{equation*}

The Dirac matrix satisfies the supersymmetry condition $\mathbf{D}_p\mathbf{Q}_p = -\mathbf{Q}_p\mathbf{D}_p$. That also means that the anti-commutator between the Dirac matrix $\mathbf{D}_p$ and block diagonal matrix $\mathbf{Q}_p$ vanishes. Further,
\begin{align}\label{eqn:conjugate}
\mathbf{D}_pv=\lambda v &\iff \mathbf{Q}_p\mathbf{D}_p\mathbf{Q}_p v = -\mathbf{D}_pv =  -\lambda v \nonumber\\
&\iff -\mathbf{Q}_p\mathbf{D}_pv =  -\lambda \mathbf{Q}_p v\nonumber\\
&\iff \mathbf{D}_p\mathbf{Q}_pv =  -\lambda \mathbf{Q}_p v,
\end{align}
which implies that $\mathbf{Q}_pv$ is an eigenvector associated with the eigenvalue $-\lambda$.

Essentially, the above shows that the Dirac operator of a simplicial complex satisfies $\mathbf{D}_pv=\lambda v$ where $\lambda$ is the eigenvalue associated with the eigenvector $v$ if and only if
\[
\mathbf{D}_p(\mathbf{Q}_pv) = -\lambda(\mathbf{Q}_pv),
\]
where $\mathbf{Q}_pv$ is the eigenvector associated to the eigenvalue $-\lambda$.

Since $\lambda$ (resp. $-\lambda$) is an eigenvalue of $\mathbf{D}_p$ with corresponding eigenvector $v$ (resp. $\mathbf{Q}_pv$), then for any positive integer $s$, $\lambda^s$ (resp. $(-\lambda)^s$) is an eigenvalue of $\mathbf{D}_p^s$ with corresponding eigenvector $v$ (resp. $\mathbf{Q}_pv$). The detailed proof is in Appendix B.

Now, we consider the relationship  between the eigenspectrum of $\mathbf{D}_p^2$ and $\mathbf{D}_p$. For the case of zero eigenvalues, $\mathbf{D}_p^2v = 0$ naturally implies $\mathbf{D}_pv=0$. Hence, $\mathbf{D}_p$ shares the same eigenvectors as $\mathbf{D}_p^2$ for zero eigenvalues. If $\lambda^2$ is a non-zero eigenvalue of $\mathbf{D}_p^2$ with eigenvector $v$, then we have the following possible cases for $\mathbf{D}_p$:
\begin{itemize}
	\item [(i)] $\lambda$ is an eigenvalue of $\mathbf{D}_p$ with eigenvector $w=(\mathbf{D}_p+\lambda I)v$. i.e. $(\mathbf{D}_p-\lambda I)w=0$.
	\item [(ii)] $-\lambda$ is an eigenvalue of $\mathbf{D}_p$ with eigenvector $w=(\mathbf{D}_p-\lambda I)v$. i.e. $(\mathbf{D}_p+\lambda I)w=0$.
\end{itemize}
It is easy to derive the above cases by considering $(\mathbf{D}_p^2-\lambda^2 I)v = 0$. Then
\begin{equation}\label{eqn:eigenvec}
	(\mathbf{D}_p-\lambda I)(\mathbf{D}_p+\lambda I)v =0.
\end{equation}
Here, there are two possible cases since by \eqref{eqn:conjugate}, either $\lambda$ or $-\lambda$ is the eigenvalue of $\mathbf{D}_p$. If $-\lambda$ is the eigenvalue of $\mathbf{D}_p$, then $(\mathbf{D}_p+\lambda I)w=0$ for some non-zero eigenvector $w$. This implies that $(\mathbf{D}_p-\lambda I)w\neq 0$, otherwise it contradicts $(\mathbf{D}_p+\lambda I)w=0$. Hence, this means that Eq. \eqref{eqn:eigenvec} can be rewritten as
\begin{equation*}
	(\mathbf{D}_p+\lambda I)w=0,
\end{equation*}
where $w=(\mathbf{D}_p-\lambda I)v$ is a non-zero eigenvector for $\mathbf{D}_p$ with corresponding eigenvalue $-\lambda$.

Similarly, if $\lambda$ is an eigenvalue of $\mathbf{D}_p$, then $(\mathbf{D}_p-\lambda I)w=0$ for some non-zero eigenvector $w$. This implies that $(\mathbf{D}_p+\lambda I)w\neq 0$, otherwise it contradicts $(\mathbf{D}_p-\lambda I)w=0$. Therefore, Eq. \eqref{eqn:eigenvec} can be rewritten as
\begin{equation*}
(\mathbf{D}_p-\lambda I)w=0,
\end{equation*}
where $w=(\mathbf{D}_p+\lambda I)v$ is a non-zero eigenvector for $\mathbf{D}_p$ with corresponding eigenvalue $\lambda$.

This leads us to the following relations connecting $\mathbf{D}_p$, $\mathbf{D}_p^2$ and $\mathbf{L}_k$ ($0\leq k \leq p+1$). For any $v\in \ker \mathbf{D}_p^2$,
\begin{align*}
	\mathbf{D}_p^2 v = \bzero &\iff
	\begin{cases}
		\mathbf{L}_0\mathbf{w}_0 = \bzero, & k=0\\
		\mathbf{L}_k\mathbf{w}_k = \bzero, & 0<k<p+1\\
		\mathbf{L}_{p+1}^{\text{down}}\mathbf{w}_{p+1} = \bzero, & k = p+1
	\end{cases},
\end{align*}
where $v = (\mathbf{w}_0^\top, \mathbf{w}_1^\top,\cdots,\mathbf{w}_{k-1}^\top,\mathbf{w}_k^\top,\mathbf{w}_{k+1}^\top,\cdots,\mathbf{w}_p^\top,\mathbf{w}_{p+1}^\top)^\top$. In other words, $v$ is a vector consisting of block vectors $\mathbf{w}_k^\top$ for $0\leq k\leq p+1$. This means that for every $0 \leq k \leq p$, $\mathbf{w}_k^\top \in \ker \mathbf{L}_k$. In the case where $k=p+1$, $\mathbf{w}_{p+1}^\top \in \ker \mathbf{L}_{p+1}^{\text{down}}$. We have,
\[
(\mathbf{w}_0^\top, \mathbf{w}_1^\top,\cdots,\mathbf{w}_{p+1}^\top)^\top \in \ker \mathbf{L}_{p+1}^{\text{down}} \oplus \bigoplus_{k=0}^{p} \ker \mathbf{L}_k .
\]
Note that for $\mathbf{w}_{p+1}^\top$, it is the eigenvector from the kernel of $\mathbf{L}_{p+1}^{\text{down}}$.

Hence, the kernel of $\mathbf{D}_p^2$ can be decomposed into a direct sum of kernels of $\mathbf{L}_k$ from $k=0$ to $k=p+1$:
\begin{equation*}\label{eqn:dshk}
\ker \mathbf{D}_p^2 = \ker \mathbf{L}_{p+1}^{\text{down}} \oplus \bigoplus_{k=0}^{p} \ker \mathbf{L}_k.
\end{equation*}
Further, we have
\begin{align}\label{hodge-dirac}\nonumber
\ker \mathbf{D}_p = \ker \mathbf{D}_p^2 &= \ker \mathbf{L}_{p+1}^{\text{down}} \oplus \bigoplus_{k=0}^{p} \ker \mathbf{L}_k \\
&\cong \ker \mathbf{L}_{p+1}^{\text{down}} \oplus\bigoplus_{k=0}^{p} H_k,
\end{align}
where $\displaystyle\bigoplus_{k=0}^{p} H_k$ refers to the direct sum of homology groups.

Therefore, the eigenvectors of $\mathbf{D}_p$ reveal both $k$-th homology and $k$-th non-homology information within the structural data for all $0\leq k \leq p+1$. Instead of eigendecomposing HL matrices for all $0\leq k \leq p+1$, one can simply eigendecompose $\mathbf{D}_p$ to obtain all of the eigenspectrums.
As the number of zero eigenvalues of $\mathbf{L}_{p+1}^{\text{down}}$ is the $\text{rank }\mathbf{B}_{p+2}^\top$ plus the $(p+1)$-th Betti number $\beta_{p+1}$, the multiplicity of zero eigenvalues in $\mathbf{D}_p$ is the $\text{rank }\mathbf{B}_{p+2}^\top$ plus the total sum of all the Betti numbers from dimension $0$ to $p+1$. That is,

\begin{equation}\label{eqn:mult}
	\dim\ker\mathbf{D}_p = \text{rank } \mathbf{B}_{p+2}^\top + \sum_{k=0}^{p+1}\beta_k.
\end{equation}

\begin{figure*}[ht]
	\centering
	\includegraphics[width=0.8\textwidth]{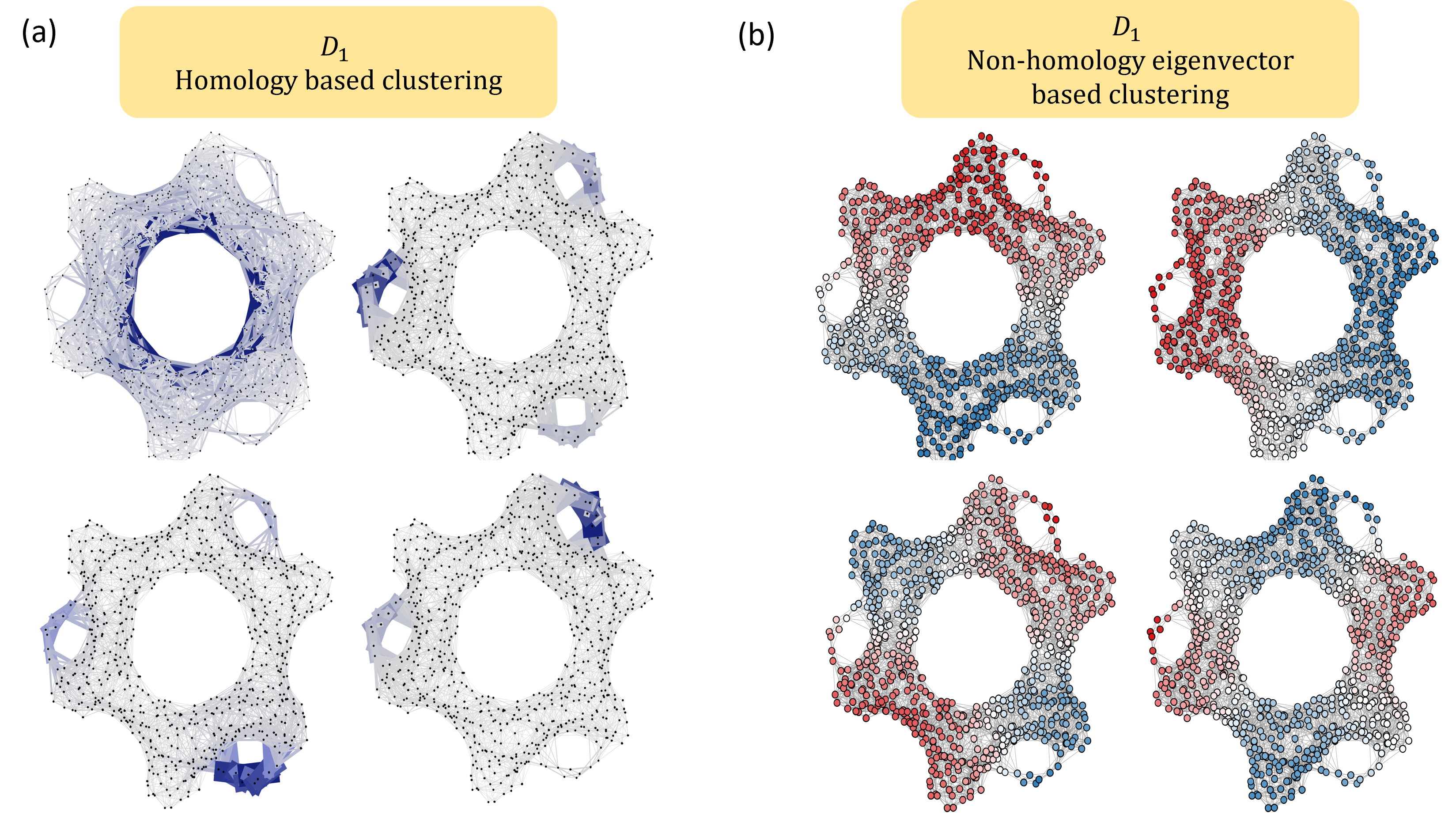}
	\caption{\label{fig:clustering1}{Illustration of loop/circle-based clustering using four one-dimensional (1D) homology generators (a)  and  spectral clustering using four zero-dimensional (0D) non-homology generators (b). The Dirac matrices $\mathbf{D}_1$ are generated from the Vietoris Rips complex of the C$_\alpha$ atoms in PDBID: 1AXC at 10\AA. (a) Here 1D homology generators $\mathbf{w}_1^\top$ are taken from the homology generators of $\mathbf{D}_1$ with eigenvalues as 0. A thick edge with dark blue color indicates large magnitude of the value, while a thinner edge with light blue color means the corresponding the 1D homology generator has a value with small magnitude on this $1$-simplex. Each 1D homology generator forms an individual loop or circle. (b) The four 0D non-homology generators $\mathbf{w}_0^\top$ are taken from the non-homology generators of $\mathbf{D}_1$ with the four smallest positive eigenvalues. Note that these 0D non-homology generators are defined on nodes (0-simplices). Nodes with negative values are colored in red while nodes with positive values are of blue color. It can be seen that the nodes in the structure can be naturally clustered into groups based on the signs of these 0D non-homology generators. }}
\end{figure*}

Mathematically, the eigenvectors corresponding to the  zero eigenvalues are known as homology generators while those from non-zero eigenvalues are the non-homology generators. Both of them can be used in structural clustering. More specifically, the homology generators can be used for clustering structures based on their loop or circle components, while non-homology generators are related to the spectral clustering, in which communities and clusters are based on their distances. Fig.~\ref{fig:clustering1} demonstrates the structural clustering with homology and non-homology generators for a protein (PDBID: 1AXC). We only consider the C$_\alpha$ atoms in structure. A Vietoris Rips complex is constructed by using a cutoff distance of 10\AA. The Dirac matrix $\mathbf{D}_1$ and its eigenvalues and eigenvectors are calculated. As the non-zero eigenvalues of $\mathbf{D}_1$ come in pairs, it suffices to consider the eigenvectors corresponding to  the positive eigenvalues. For all the non-negative eigenvalues of $\mathbf{D}_1$, the eigenvectors are arranged in ascending order according to its corresponding eigenvalues.

Fig.~\ref{fig:clustering1}(a) illustrates the loop/circle-based clustering using four one-dimensional (1D) homology generators. Note that these 1D homology generators $\mathbf{w}_1^\top$ are taken from the homology generators of $\mathbf{D}_1$ (with eigenvalues  $0$). More specifically, these 1D homology generators are defined by the $1$-simplices. In Fig.~\ref{fig:clustering1}(a), a thick edge with dark blue color indicates large magnitude of the value, while a thinner edge with light blue color means the corresponding 1D homology generator has a value with small magnitude on this $1$-simplex. It can be seen that edges with large magnitudes are the $1$-simplices that form circles or loops. Each 1D homology generator forms an individual loop or circle. In this way, 1D homology generators can be used for loop/circle-based clustering of molecular structures.

Fig.~\ref{fig:clustering1}(b) illustrates the spectral clustering using four zero-dimensional (0D) non-homology generators. The four 0D non-homology generators $\mathbf{w}_0^\top$ are taken from the non-homology generators of $\mathbf{D}_1$ with the four smallest positive eigenvalues. Note that these 0D non-homology generators are defined on nodes ($0$-simplices). In Fig.~\ref{fig:clustering1}(b), nodes with negative values are colored in red while nodes with positive values are of blue color. It can be seen that the nodes in the structure can be naturally clustered into groups based on the signs of these 0D non-homology generators. This approach is known as spectral clustering and widely used in data analysis. It should be noticed that using the higher order Dirac matrices, we can cluster not only nodes (0-simplices), but also higher dimensional simplices.

\begin{figure*}
	\centering
	\includegraphics[width=.9\textwidth]{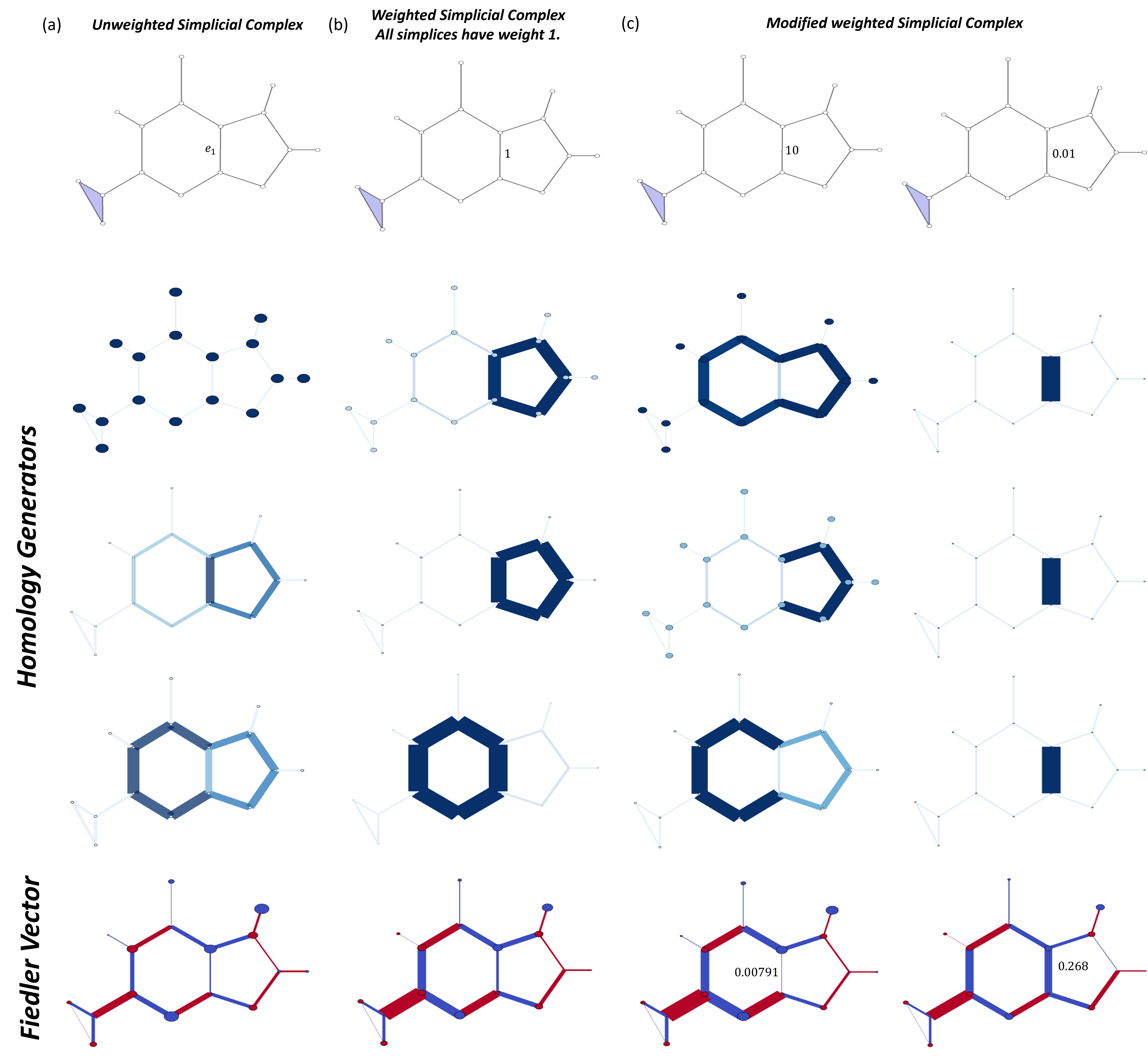}
	\caption{\label{fig:MDO_homo}Illustration of three homology generators and Fiedler vector from (a): discrete Dirac matrix and (b)-(c): weighted Dirac matrix (from weighted simplicial complexes). For the discrete Dirac matrix, the three homology generators represents one 1D component and two 2D circles. By assigning simplex $\sigma$ with different weight $w_\sigma$s, three weighted simplicial complexes are constructed in (b) and (c). In Fig. \ref{fig:MDO_homo}(b), the weighted simplicial complex consists of all weights $w_\sigma$ equal to $1$. Fig. \ref{fig:MDO_homo}(c) shows two weighted simplicial complexes by changing the weights of edge $e_1$ from 1 to $10$ and $0.01$ while the rest of weights remain unchanged. The magnitude of the homology generators are influenced by these weights and are reflected based on their thickness and darkness. For the homology generators, the edges (or vertices) are thicker and in darker blue color if they have a larger magnitude. Similarly, the edges and vertices are colored in red/blue if their elements in the Fiedler vectors have positive/negative sign. The magnitudes of their values in Fiedler vectors are represented by the thickness of edges and size of vertices.
	}
\end{figure*}

\paragraph{Spectral of weighted Dirac matrix}
The weighted Dirac matrix has different spectral properties based on the different weighting schemes. Fig. \ref{fig:MDO_homo} illustrates the spectrum  of the weighted Dirac matrix defined from the guanine molecule structure (using all-atom representation). We construct an unweighted Vietoris Rips complex using a cutoff distance of 1.2\AA. The discrete Dirac matrix $\mathbf{D}_1$ can be computed using \eqref{eqn:dirac}. The discrete Dirac matrix $\mathbf{D}_1$ is eigendecomposed to obtain its eigenvalues and eigenvectors. Moreover, a weighted simplicial complex is constructed by assigning  simplex $\sigma$ with different weight $w_\sigma$. The metric matrices $\mathbf{G}_p$ are computed and weighted Dirac matrix $\overline{\mathbf{D}}_1$ can then be constructed. Fig. \ref{fig:MDO_homo} shows the homology generators {and Fiedler vector} for an unweighted simplicial complex and three different weighted simplicial complexes. Among the three weighted simplicial complexes, Fig. \ref{fig:MDO_homo}(b) shows a weighted simplicial complex where all weights $w_\sigma$ are equal to $1$. Two modified weighted simplicial complexes are constructed by modifying the weights of edge $e_1$ ranging from $10$ and $0.01$ with all the other weights kept unchanged. With the same underlying simplicial complex, they share the same three homology generators, one 1D component and two 2D circles.  Fig.~\ref{fig:MDO_homo} shows the corresponding eigenvectors for these homology generators.  The magnitude of the eigenvectors are represented by the thickness and darkness. An edge (or vertex) with thicker lines and darker blue color indicates a larger magnitude.

In general, the weight of a simplex has an inverse effect on the corresponding element  of the homology eigenvectors (i.e., homology generators). When the simplex has a smaller weight, the corresponding element of the homology eigenvectors has  larger magnitude. Similar patterns also appear in non-homology generators. Fig.~\ref{fig:MDO_homo} illustrates the Fiedler vectors (i.e., eigenvector corresponding to the first smallest non-zero eigenvalue) of the nonweighted and weighted simplicial complexes. Simplices are colored in red/blue if the element of the  non-homology eigenvectors has value positive/negative. The thickness of simplices represents the magnitude of their values in non-homology generators. It can be seen clearly that the weight of a simplex has an inverse effect on its magnitude of the values of eigenvectors.


\subsection{Persistent Dirac}%

\paragraph{Mathematical foundation for Persistent Dirac analysis}

Recently, persistent Laplacian and persistent sheaf Laplacians have been developed \cite{memoli2020persistent, wei2021persistent}. Their essential idea is to explore the persistence of spectral information during the filtration process. Here we develop the rigorous mathematical framework for persistent Dirac.

Let $(\R, \leq)$ be a category of real numbers with morphisms given by $a\rightarrow b$ for any $a\leq b$. A functor $\mathcal{F}:(\R, \leq) \rightarrow \textbf{Simp}$ gives a filtration of simplicial complexes of finite type, i.e. $\mathcal{F}$ maps from a category of real numbers to a category of simplicial complexes of finite type. For any two real numbers $a\leq b$, the functor $\mathcal{F}$ satisfies the inclusion
\[
\mathcal{F}(a) \hookrightarrow \mathcal{F}(b),
\]
which induces a morphism of chain complexes
\[
C_*(\mathcal{F}(a), \R) \hookrightarrow C_*(\mathcal{F}(b), \R).
\]
Let $\mathcal{F}(\infty)=\displaystyle\bigcup_{a\in \R} \mathcal{F}(a)$ and $C_*=C_*(\mathcal{F}(\infty), \R)$. Note that $C_*$ can be endowed with an innerproduct $\langle\cdot,\cdot\rangle$. Further, a subspace $C_*(\mathcal{F}(a), \R)$ would inherit the inner product structure of $C_*$ and a boundary operator given by the restriction
\[
\partial^a_p = \partial_p|_{C_p(\mathcal{F}(a),\R)}: C_p(\mathcal{F}(a), \R) \rightarrow C_{p-1}(\mathcal{F}(a), \R).
\]

Here $\partial_*$ is the boundary operator of $C_*$. For convenience, we shall write $C^a_p = C_p(\mathcal{F}(a), \R)$. For a pair of simplicial complexes $\mathcal{F}(a) \subset \mathcal{F}(b)$, we consider the inclusion map  $\iota:\mathcal{F}(a)\hookrightarrow \mathcal{F}(b)$. For $p\in\N$, the subspace
\[
C_p^{a,b} := \{x\in C^b_p: \partial_p^b(x)\in C^a_{p-1}\} \subseteq C^b_p,
\]
which consists of the $p$-chains in $C^b_p$ such that their images are under the boundary operator $\partial_p^b$ in the subspace $C^a_{p-1}$ of $C^b_{p-1}$. Also, we have a linear operator
\[
\partial_{p}^{a,b} = \partial_p^b|_{C_p^{a,b}} : C_p^{a,b} \rightarrow C_{p-1}^{a},
\]
which induces an adjoint operator
\[
(\partial_{p}^{a,b})^*: C_{p-1}^a \rightarrow C_p^{a,b}
\]
with respect to the inner product $\langle\cdot,\cdot\rangle$.

Let $n^{a,b}_p := \dim (C^{a,b}_p)$. Then following commutative diagram is thus induced by $\iota$.

\begin{figure}[H]
	\[\begin{tikzcd}
	\cdots & {C_{p-1}^a} & {C_{p}^a} & {C_{p+1}^a} & \cdots \\
	\cdots & {C_{p-1}^b} & {C_{p}^b} & {C_{p+1}^b} & \cdots
	\arrow["{\partial^a}",from=1-2, to=1-3]
	\arrow["{\partial^a}",from=1-3, to=1-4]
	\arrow["{\iota}",dashed, hook, from=1-2, to=2-2]
	\arrow["{\iota}",dashed, hook, from=1-3, to=2-3]
	\arrow["{\iota}",dashed, hook, from=1-4, to=2-4]
	\arrow["{\partial^a}",from=1-1, to=1-2]
	\arrow["{\partial^b}",from=2-1, to=2-2]
	\arrow["{\partial^b}",from=2-2, to=2-3]
	\arrow["{\partial^b}",from=2-3, to=2-4]
	\arrow["{\partial^a}",from=1-4, to=1-5]
	\arrow["{\partial^b}",from=2-4, to=2-5]
	\end{tikzcd}\]
\end{figure}

Notice that $\partial_p^{a,b}$ is a restriction to $C_p^{a,b}$ in order to obtain the ``diagonal" operators $\partial_p^{a,b} : C_p^{a,b} \rightarrow C_{p-1}^a$. Similarly, with a restriction to $C_p^{a,b}$, we can then define the $p$-dimensional boundary matrices $\mathbf{B}_{p}^{a,b}$ which consists of every entry value of $\partial_p^{a,b}$.

The persistent Dirac operator $\mathbf{D}_{p}^{a,b}$ can then be written as follows.
\begin{widetext}
\begin{equation*}
	\renewcommand{\arraystretch}{1.5}
\mathbf{D}_{p}^{a,b}=\begin{bmatrix}
	\bzero_{n_0\times n_0}    & \mathbf{B}_{1}^{a,b}    & \bzero_{n_0\times n_2}     & \cdots      & \bzero_{n_0\times n_{p}} & \bzero_{n_0\times n_{p+1}}  \\
	
	(\mathbf{B}_{1}^{a,b})^\top  & \bzero_{n_1\times n_1}      & \mathbf{B}_{2}^{a,b}    & \cdots      & \bzero_{n_1\times n_{p}} & \bzero_{n_1\times n_{p+1}}  \\
	
	\bzero_{n_2\times n_0}      & (\mathbf{B}_{2}^{a,b})^\top  & \bzero_{n_2\times n_2}       & \cdots    & \bzero_{n_2\times n_{p}} & \bzero_{n_2\times n_{p+1}}   \\
	
	\vdots & \vdots & \vdots & \vdots & \vdots & \vdots  \\
	
	\bzero_{n_{p}\times n_0}       & \bzero_{n_{p}\times n_1}       & \bzero_{n_{p}\times n_2}       & \cdots       &  \bzero_{n_{p} \times n_{p}}       & \mathbf{B}_{p}^{a,b}\\
	
	\bzero_{n_{p+1}\times n_0}       & \bzero_{n_{p+1}\times n_1}       & \bzero_{n_{p+1}\times n_2}       & \cdots       & (\mathbf{B}_{p}^{a,b})^\top  & \bzero_{n_{p+1}\times n_{p+1}}
	\end{bmatrix}.
\end{equation*}
\end{widetext}
The maps and spaces are also illustrated in the diagram below
\begin{figure}[H]
	\[\begin{tikzcd}
	C_{p+1}^a \arrow["{\iota}",dd, dashed, hook] \arrow[rrrr, "\partial_{p+1}^a"] &  &                                                                             &  & C_p^a \arrow[lld, "{(\partial_{p+1}^{a,b})^*}", shift left=2] \arrow["{\iota}", dd, dashed, hook, shift left] \arrow[rr, "\partial_p^a", shift left] &  & C_{p-1}^a \arrow["{\iota}",dd, dashed, hook] \arrow[ll, "(\partial_p^a)^*", shift left] \\
	&  & {C_{p+1}^{a,b}} \arrow[,lld, dashed, hook] \arrow[rru, "{\partial_{p+1}^{a,b}}"] &  &                                                                                                                                       &  &                                                                               \\
	C_{p+1}^b \arrow[rrrr, "\partial_{p+1}^b"]                          &  &                                                                             &  & C_p^b \arrow[rr, "\partial_{p}^b"]                                                                                                    &  & C_{p-1}^b
	\end{tikzcd}\]
\end{figure}
Further, the $p$-th persistent Hodge Laplacian can be defined as
\[
\mathbf{L}_p^{a,b} = \begin{cases}
\mathbf{B}_{1}^{a,b}(\mathbf{B}_{1}^{a,b})^\top, & p =0 \\
(\mathbf{B}_p^{a,b})^\top\mathbf{B}_p^{a,b} + \mathbf{B}_{p+1}^{a,b}(\mathbf{B}_{p+1}^{a,b})^\top, & p>0.
\end{cases}
\]

Similarly, the matrices $(\mathbf{B}_p^{a,b})^\top\mathbf{B}_p^{a,b}$ and $\mathbf{B}_{p+1}^{a,b}(\mathbf{B}_{p+1}^{a,b})^\top$ are the $p$-th persistent lower and upper Hodge Laplacians $(\mathbf{L}_{p+1}^{\text{down}})^{a,b}$ and $(\mathbf{L}_{p+1}^{\text{up}})^{a,b}$ respectively.
Based on \eqref{hodge-dirac}, the following result shows that the nullity of $p$-th persistent Dirac operator equals to the rank of $\mathbf{B}_{p+2}^\top$ plus the sum of $k$-th persistent Betti numbers, where $0\leq k \leq p+1$.
\begin{align}\label{eqn:pers_hodge_dirac} \nonumber
	\ker \mathbf{D}_p^{a,b} = \ker (\mathbf{D}_p^{a,b})^2 &= \ker (\mathbf{L}_{p+1}^{\text{down}})^{a,b} \oplus\bigoplus_{k=0}^{p} \ker \mathbf{L}_k^{a,b} \\\nonumber
	&\cong \ker (\mathbf{L}_{p+1}^{\text{down}})^{a,b} \oplus\bigoplus_{k=0}^{p} (H_k)^{a,b},
\end{align}
where $\displaystyle\bigoplus_{k=0}^{p} (H_k)^{a,b}$ refers to the direct sum of $(a,b)$-persistent homology groups. The $(a,b)$-persistent homology groups characterizes the homology generators that are born at time $a$ and survive to time $b$.

\begin{figure*}
	\centering
	\includegraphics[width=\textwidth]{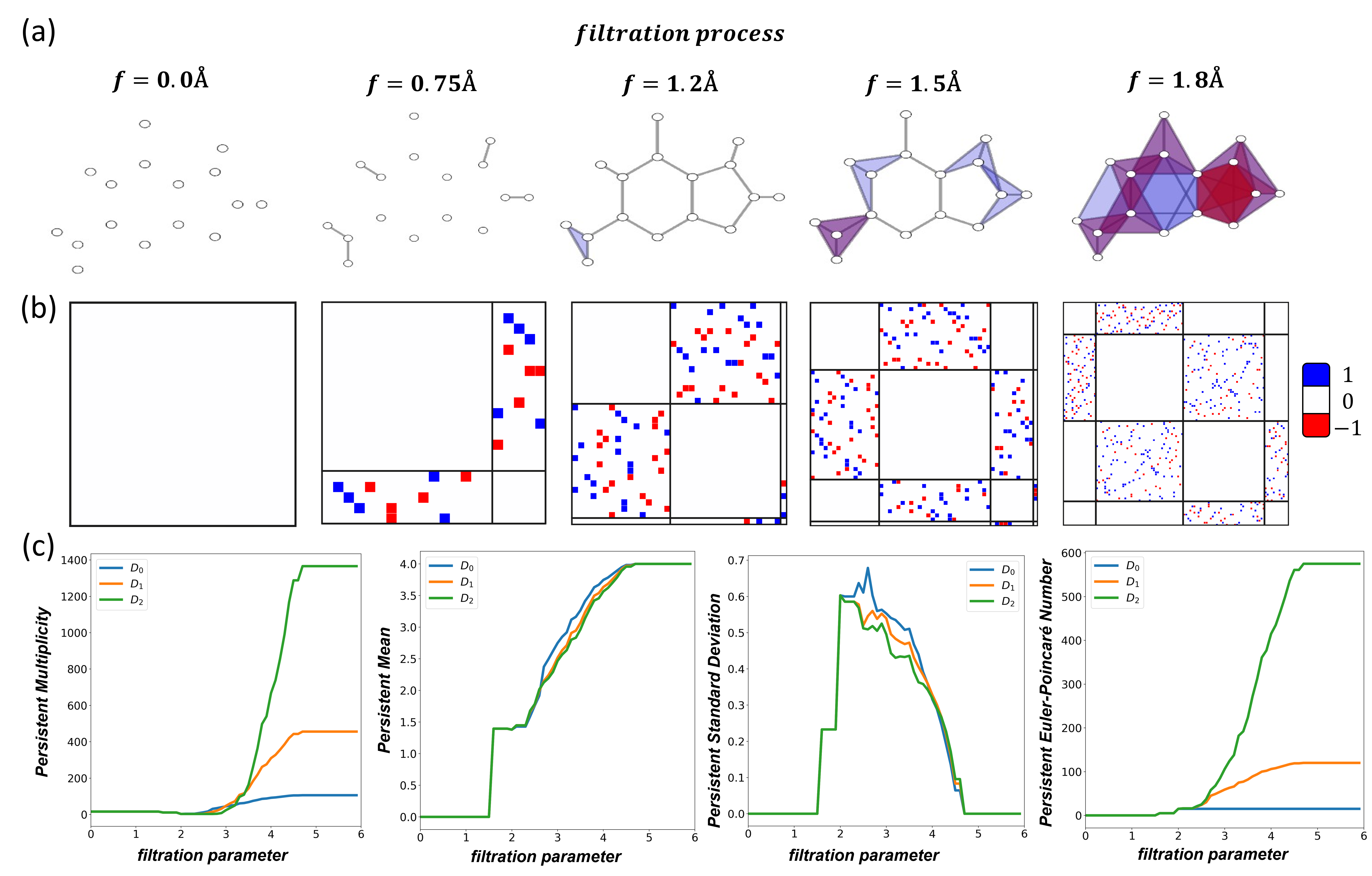}
	\caption{\label{fig:filtration}Illustration of  the filtration process of the guanine molecule (a), its associated Dirac matrices (b), and persistent attributes (c). In the filtration process, more simplices are formed in simplicial complex and the size of Dirac matrix increases. The eigenspectrum of Dirac matrices changes in the filtration process. The changes in eigenspectrum are being converted into a series of 12 statistical and combinatorial attributes. One of the statistical attribute, persistent multiplicity, provides quantitative analysis to the change in zero eigenvalues of Dirac matrices while the remaining 11 persistent attributes are derived from the non-zero eigenvalues.}
\end{figure*}

Fig. \ref{fig:filtration} illustrates the persistent Dirac analysis of the guanine molecule (using all-atom representation). More specifically, Fig.~\ref{fig:filtration}(a) shows the Vietoris-Rips complex of the guanine molecule  when filtration parameter $f= 0.0$\AA, $0.75$\AA, $1.2$\AA, $1.5$\AA\space and $1.8$\AA. In particular, triangles first appear around 1.2\AA \space and tetrahedron starts to appear at 1.5\AA. Fig.~\ref{fig:filtration}(b) shows the corresponding Dirac matrix $\mathbf{D}_2$. The size of the Dirac matrix $\mathbf{D}_2$ consistently increases during the filtration process.

\paragraph{Persistent attributes}

For any Dirac matrix, its non-zero eigenvalues come in pairs. Each pair contains one negative eigenvalue and one positive counterpart. For the set of all its positive eigenvalues, a Dirac Zeta function can be defined as follows \cite{knill2013dirac},
\[
\zeta(s) = \sum_{j=1}^n \frac{1}{\lambda_j^s} = \sum_{j=1}^n e^{-s\log \lambda_j}, s\in \mathbb{C}.
\]
Here $\zeta(-m) = \sum_{i=1}^n \lambda_i^m$, $m\in \mathbb{Z}$ is the $m$-th spectral moments of DO matrices and $\zeta(-1)$ is the Laplacian graph energy. Another way to define Dirac Zeta function is to consider its negative eigenvalues by replacing the $\lambda_j^{-s}$ with $(1+e^{-i\pi s})|\lambda_j|^{-s}$. Here $\lambda_j$ can be negative. For instance, $\zeta(2) = 2\sum_{j=1} \lambda_j^{-2}$.

Furthermore, the $q$-Dirac complexity of a simplicial complex $\mathcal{K}$ can be defined as
\[
c_q(\mathbf{D}_p) = \prod_{\substack{\lambda_j\neq0\\\lambda_j\in\sigma(\mathbf{D}_p)}} \lambda_j^q.
\]

The case where $q=1$ is introduced in \cite{knill2013mckean}. $c_1(\mathbf{D}_p)$ is equal to the product of all non-zero eigenvalues in spectra of $\mathbf{D}_p$ since the non-zero eigenvalues come in pairs. The number of non-zero eigenvalues pairs in $\mathbf{D}_p$ is the (signless) Euler-Poincar\'{e} number defined as follows,
$$\ell = \frac{1}{2}\sum_{k=0}^{p+1} n_k-\frac{1}{2}\dim\ker\mathbf{D}_p$$
where $n_k$ is the number of $k$-simplices and $\dim\ker\mathbf{D}_p$ is the multiplicity of zero eigenvalues of $\mathbf{D}_p$.

Using Eq. \eqref{eqn:mult}, $\ell$ can be computed as follows:
\begin{equation}\label{eqn:euler-poincare}
	\ell = \frac{1}{2}\sum_{k=0}^{p+1} (n_k - \beta_k) - \frac{1}{2}\text{rank } \mathbf{B}_{p+2}^\top.
\end{equation}

Interestingly, the spanning tree number, introduced as one of the spectral indices in molecular descriptors \cite{puzyn2010recent}, can be written as
\[
t(\mathbf{D}_p) = \frac{1}{2}\log(c_1(\mathbf{D}_p)) - \log(\ell+1),
\]
Alternatively,
$t(\mathbf{D}_p) = \displaystyle\log\left[\frac{1}{\ell+1}\cdot\sqrt{c_1(\mathbf{D}_p)}\right]$.

To summarize, we consolidate and consider a set of statistical and combinatorial attributes as molecular descriptors for each given set of positive eigenvalues $\{\lambda_1, \lambda_2, \cdots, \lambda_\ell\}$ where $\ell$ is the number of non-zero eigenvalue pairs:
\begin{itemize}
	\item $\min\{\lambda_1, \lambda_2, \cdots, \lambda_k\}$, also known as the Fiedler value.
	\item $\max\{\lambda_1, \lambda_2, \cdots, \lambda_n\}$
	\item $\bar{\lambda} = \frac{1}{n}\sum_{i=1}^n \lambda_i = \frac{1}{n}\zeta(-1)$.
	\item Standard Deviation
	\item Laplacian Graph Energy $\zeta(-1)$.
	\item (Signless) Euler-Poincar\'{e} Number (number of non-zero eigenvalue pairs) $\ell$
	\item Generalised Mean Graph Energy $\sum_{i=1}^n \frac{|\lambda_i-\bar{\lambda}|}{n}$.
	\item Spectral 2nd Moment $\zeta(-2)$.
	\item $\zeta(2) = 2\sum_{j=1}^n \lambda_j^{-2}$.
	\item Quasi-Wiener Index $(n+1)\zeta(1)$.
	\item Spanning Tree Number $t(\mathbf{D}_p)$.
\end{itemize}

Fig.~\ref{fig:filtration}(c) shows the persistent multiplicity, persistent mean, persistent standard deviation and persistent (signless) Euler-Poincar\'{e} number for the filtration of guanine molecule. Further information such as the persistent multiplicities of $\mathbf{L}_k$ ($0\leq k\leq 2)$ and $\mathbf{L}_k^{\text{down}}$ ($1\leq k \leq 3$) can be found in Appendix F.
Recall that the persistent multiplicity is equivalent to the persistent Betti number. Here, the persistent multiplicity and persistent (signless) Euler-Poincar\'{e} number of $\mathbf{D}_p$ can be quantitatively analysed by comparing the persistent multiplicity of $\mathbf{L}_{p+1}^{\text{down}}$ and the $k$-th persistent Betti numbers for $0\leq k \leq p$. It can be seen that these persistent attributes change with the filtration value. Each variation of the persistent attribute indicates a certain change in the simplicial complex.

At the very start of the filtration, there are 16 isolated atoms which means that there are 16 connected components. Hence, the persistent multiplicity of $\mathbf{L}_0$ is 16 since $\beta_0 = 16$. As all other Betti numbers are zero and there are no higher order simplices present at the start of the filtration, $\mathbf{D}_0$, $\mathbf{D}_1$ and $\mathbf{D}_2$ are all-zero $16\times16$ matrices. Therefore, the persistent multiplicity of $\mathbf{D}_0$, $\mathbf{D}_1$ and $\mathbf{D}_2$ are all equal to 16. Using Eq. \eqref{eqn:euler-poincare}, the persistent (signless) Euler-Poincar\'{e} number is zero.

As filtration parameter $f$ increases, the size of $\mathbf{D}_0$, $\mathbf{D}_1$ and $\mathbf{D}_2$ matrix increases as well. This differs from the Hodge Laplacian matrix $\mathbf{L}_0$, whose size remains unchanged.

At filtration size 4.7\AA, a complete simplicial complex is achieved, i.e., any $p+1$ vertices will form a $p$-simplex. When this happens, the size of $\mathbf{D}_p$ no longer increases any further. Here, the size of $\mathbf{D}_0$, $\mathbf{D}_1$ and $\mathbf{D}_2$ are distinct. The size of $\mathbf{D}_0$ is $136\times136$ since $\frac{16\times 15}{2}$ (no. of $1$-simplices) + 16 (no. of $0$-simplices) = 136. Similarly, the sizes of $\mathbf{D}_1$ and $\mathbf{D}_2$ are $696\times 696$ and $2516\times2516$ respectively. Furthermore, the persistent multiplicity of $\mathbf{D}_0$, $\mathbf{D}_1$ and $\mathbf{D}_2$ are also distinct. Using Eq. \eqref{eqn:mult}, the persistent multiplicity of $\mathbf{D}_0$ is 105 (persistent multiplicity of $\mathbf{L}_{1}^{\text{down}}$) and 1 (0-dimensional persistent Betti number) which sums up to 106. Since the persistent multiplicity of $\mathbf{L}_1$ (see Appendix F) is zero, then Eq. \eqref{eq:kernel_lp_down} implies that the rank of $\mathbf{B}_2^\top$ is 105. In addition, the persistent multiplicity of $\mathbf{D}_1$ and $\mathbf{D}_2$ are 456 and 1366 respectively. Based on the non-zero eigenvalues, the persistent (signless) Euler-Poincar\'{e} number of $\mathbf{D}_0$, $\mathbf{D}_1$ and $\mathbf{D}_2$ is 15, 120 and 575 due to Eq. \eqref{eqn:euler-poincare}.

\section{\label{sec:level4} Persistent Dirac for molecular structure representation}

\begin{figure*}
	\centering
	\includegraphics[width=0.8\textwidth]{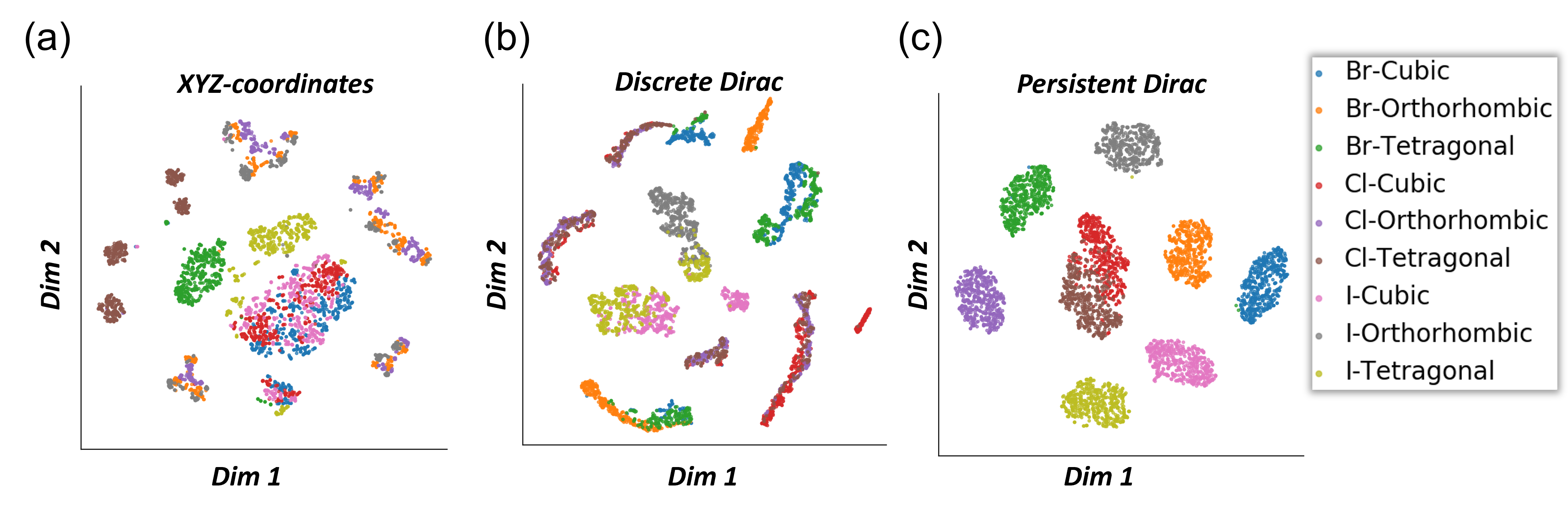}
	\caption{\label{fig:clustering}The clustering of 9 types of OIHP molecular dynamics (MD) trajectories. Three feature generation schemes are considered, including (a) XYZ-coordinates, (b) Discrete Dirac at 3.5\AA\space and (c) Persistent Dirac. Each trajectory contains 1000 configurations and $t$-SNE model is used for clustering (of the last 500 configurations at equilibrium). The x-axis and y-axis are the two principal components obtained from the $t$-SNE model.}
\end{figure*}

Recently, a series of persistent models, including persistent homology, persistent spectral, persistent Ricci curvature, and persistent Laplacian, have demonstrated their great power in molecular representations \cite{wee2021ollivier,meng2021persistent,wee2021forman,wee2022persistent}. They have consistently outperformed traditional graph-based models in various tasks of drug design. Here we study the representation capability of Persistent Dirac in molecular data analysis.

We consider the Organic-inorganic halide perovskite (OIHP) dataset. More specifically, three kinds of Methylammonium lead halides (MAPbX$_3$, X$=$Cl, Br, I), i.e., orthorhombic, tetragonal, and cubic phase of MAPbX$_3$ are used. For each kind, there are 3 types of X atoms, including chlorine Cl, bromine Br and iodine I. The molecular dynamic simulations are systematically carried out on these molecular structures with the initial configurations based on pre-defined crystal cell parameters. For each MAPbX$_3$ structure, 1000 configurations are equally sampled from its MD simulation trajectory and the last 500 configurations, which represent stable structures, are selected for the test of our persistent Dirac model. Essentially, a total of 4500 configurations from the 9 types of MAPbX$_3$ structures are mixed together and our persistent Dirac based molecular fingerprint is used in the clustering of these configurations.

Computationally, our persistent Dirac is generated based on Alpha complex and the filtration parameter is the distance. More specifically, for each frame, an Alpha complex is constructed from its coordinate data and applied in a filtration process. The Dirac matrices $\mathbf{D}_0$ and $\mathbf{D}_1$ are computed from 1\AA\space to 6.5\AA\space with stepsize 0.25\AA\space throughout the filtration process. Hence, the eigenvalues of $\mathbf{D}_0$ and $\mathbf{D}_1$ each contribute to 12 statistical attributes for 23 timesteps per frame. The feature size sums up to 552. By considering with and without hydrogen atoms, the total feature size for persistent Dirac is $552\times 2 = 1104$. Likewise, for coordinate-only model, the input features are $xyz$-coordinates of all the atoms. Since each structure consists of 553 atoms, the feature size is of $553\times3=1659$. For the discrete Dirac model, the feature size is $552$. The clustering of these MAPbX$_3$ structures is then studied using unsupervised learning models, in particular $t$-distributed stochastic neighbor embedding ($t$-SNE).

Fig.~\ref{fig:clustering} illustrates the comparison of the clustering results from three different models, including coordinate-only model (xyz-coordinate) (a), discrete Dirac (b), and persistent Dirac (c). It can be seen that our persistent Dirac model demonstrates better capabilities in characterizing the intrinsic structure information and discriminating the 9 types of OIHPs clearly. In our persistent Dirac model, the filtration process at various scales provided the geometrical information needed to balance the topological information. The combination of topological and geometrical information contributes to the success of our persistent Dirac model in OIHP clustering. Fig.~\ref{fig:clustering}(b) shows the performance of Dirac matrix related statistical attributes at filtration value 3.5\AA. Even though it shows certain clustering effects, the overall performance is not as good as persistent Dirac. Additional clustering tests are performed for discrete Dirac model at 3\AA\space and 4\AA\space in Appendix E. Similarly, statistical attributes of discrete Dirac model at a single scale fail to distinguish the 9 types of OIHPs.

\section{Conclusion}

Molecular representations are essential to the modeling and analysis of molecular systems. Motivated by the great success of persistent Laplacian, we develop the first persistent Dirac-based molecular representation and fingerprint. A rigorous theoretical framework for persistent Dirac is introduced through the commutative diagram of discrete Dirac operator over a filtration process. Moreover, a series of persistent attributes, which characterize the persistence and variations of the eigenspectrum of Dirac matrices, are proposed and further used as molecular fingerprints. The eigenspectrum properties of discrete Dirac matrices have been studied, in particular, the geometric and topological properties of both non-homology and homology eigenvectors. We also consider weighted Dirac model and the influence of weighting schemes on eigenspectrum information. Finally, our persistent Dirac-based models have been used in the clustering of molecular configurations from nine types of organic-inorganic halide perovskites.
This work could open new perspectives for the use of persistent Dirac-based molecular fingerprints. We hope that this can inspire future interdisciplinary work between Dirac operators and machine learning along OIHPs or other relevant research directions. An interesting direction for further exploration would be the use of non-symmetric persistent Dirac features in predicting biological, chemical and physical properties in biomolecular data. For instance, further exploration in the use of non-symmetric persistent Dirac features can be considered in the prediction of energy bandgap and other material properties in OIHPs \cite{anand2022topological}.

\bibliography{refs}

\clearpage

\newpage

\setcounter{figure}{0}
\renewcommand{\thefigure}{S\arabic{figure}}
\section*[Appendix]{Appendix A: Proofs of Elementary Properties of Upper and Lower Hodge Laplacians}
\begin{proof}
	\begin{itemize}
		\item[(i)] For any $v\in \ker \overline{\mathbf{L}}_p^{\text{down}}$, $v$ satisfies $\overline{\mathbf{L}}_p^{\text{down}}v=\bzero$, then $v^\top\overline{\mathbf{B}}_p^\top\overline{\mathbf{B}}_pv = v^\top\overline{\mathbf{L}}_p^{\text{down}}v=\bzero$ which shows that $\overline{\mathbf{B}}_p v=\bzero$. Hence, $v\in \ker \overline{\mathbf{B}}_p$.
		On the other hand, for any $v\in \ker \overline{\mathbf{B}}_p$, we have $\overline{\mathbf{B}}_p v=\bzero$. Multiplying both sides by $\overline{\mathbf{B}}_p^\top$ implies that $\overline{\mathbf{B}}_p^\top\overline{\mathbf{B}}_pv =\bzero$.
		
		\item[(ii)] For any $v\in \ker \overline{\mathbf{L}}_{p-1}^{\text{up}}$, $v$ satisfies $\overline{\mathbf{L}}_{p-1}^{\text{up}}v=\bzero$, then $v^\top\overline{\mathbf{B}}_p\overline{\mathbf{B}}_p^\top v = \bzero$  which shows that $\overline{\mathbf{B}}_p^\top v=\bzero$. Hence, $v\in \ker \overline{\mathbf{B}}_p^\top$.
		On the other hand, for any $v\in \ker \overline{\mathbf{B}}_p^\top$, we have $\overline{\mathbf{B}}_p^\top v=\bzero$. Multiplying both sides by $\overline{\mathbf{B}}_p$ implies that $\overline{\mathbf{B}}_p\overline{\mathbf{B}}_p^\top v =\bzero$.
		
		\item[(iii)] Since
		\begin{align*}
			\overline{\mathbf{L}}_p^{\text{down}}v=\lambda v &\iff \overline{\mathbf{B}}_p\overline{\mathbf{B}}_p^\top\overline{\mathbf{B}}_pv=\lambda \overline{\mathbf{B}}_pv\\
			&\iff \overline{\mathbf{L}}_{p-1}^{\text{up}}\overline{\mathbf{B}}_pv=\lambda \overline{\mathbf{B}}_pv,
		\end{align*}
		then $\lambda$ is a non-zero eigenvalue of $\overline{\mathbf{L}}_{p}^{\text{down}}$ with corresponding eigenvector $\overline{\mathbf{B}}_pv$.
		
		\item[(iv)] Similar to (iii), $\overline{\mathbf{L}}_p^{\text{down}}v=0 \iff \overline{\mathbf{L}}_{p-1}^{\text{up}}\overline{\mathbf{B}}_pv =0$.
		\item[(v)] For any $v \in \operatorname{im} \overline{\mathbf{L}}_p^{\text{up}}$, there exist some $w$ such that $\overline{\mathbf{L}}_p^{\text{up}}w = v$. Hence,
		\begin{align*}
			\overline{\mathbf{L}}_p^{\text{down}} v &= \overline{\mathbf{B}}_p^\top\overline{\mathbf{B}}_pv = \overline{\mathbf{B}}_p^\top\overline{\mathbf{B}}_p\overline{\mathbf{L}}_p^{\text{up}}w \\
			&= \overline{\mathbf{B}}_p^\top\underbrace{\overline{\mathbf{B}}_p\overline{\mathbf{B}}_{p+1}}_{=\bzero}\overline{\mathbf{B}}_{p+1}^\top w= \bzero.
		\end{align*}
		
		\item[(vi)] For any $v \in \operatorname{im} \overline{\mathbf{L}}_p^{\text{down}}$, there exist some $w$ such that $\overline{\mathbf{L}}_p^{\text{down}}w = v$. Hence,
		\begin{align*}
			\overline{\mathbf{L}}_p^{\text{up}} v &= \overline{\mathbf{B}}_{p+1}\overline{\mathbf{B}}_{p+1}^\top v = \overline{\mathbf{B}}_{p+1}\overline{\mathbf{B}}_{p+1}^\top\overline{\mathbf{L}}_p^{\text{down}}w \\
			&= \overline{\mathbf{B}}_{p+1}\underbrace{\overline{\mathbf{B}}_{p+1}^\top\overline{\mathbf{B}}_{p}^\top}_{=\bzero}\overline{\mathbf{B}}_pw= \bzero.
		\end{align*}
		
		\item[(v)] Define the orthogonal complement
		\[
		\ker(\overline{\mathbf{B}}_p)^\perp = \{c \in C_p|c\perp d, \quad \forall d \in \ker(\overline{\mathbf{B}}_p)\}.
		\]
		Note that $\ker(\overline{\mathbf{B}}_p)^\perp = \operatorname{im}(\overline{\mathbf{B}}_p^\top)$. This is easily seen since for any $c \in \ker(\overline{\mathbf{B}}_p)^\perp$, we have $\overline{\mathbf{B}}_p(c) = d \neq 0$ and hence
		\[
		\overline{\mathbf{B}}_p^\top\overline{\mathbf{B}}_p(c)= \overline{\mathbf{B}}_p^\top(d) = c.
		\]
		Similarly, $\overline{\mathbf{B}}_p(c) = \overline{\mathbf{B}}_p\overline{\mathbf{B}}_p^\top(d) = d$.
		
		By replacing $\overline{\mathbf{B}}_p$ with $\overline{\mathbf{B}}_p^\top$ and $\overline{\mathbf{B}}_p^\top$ with $\overline{\mathbf{B}}_p$, one obtains
		\[
		\ker(\overline{\mathbf{B}}_p^\top)^\perp = \operatorname{im}(\overline{\mathbf{B}}_p).
		\]
		Lastly, by taking orthogonal complement on both sides,
		\[
		\ker \overline{\mathbf{B}}_p^\top = (\operatorname{im } \overline{\mathbf{B}}_p)^\perp.
		\]
	\end{itemize}
\end{proof}

\begin{figure*}
	\centering
	\includegraphics[width=0.8\textwidth]{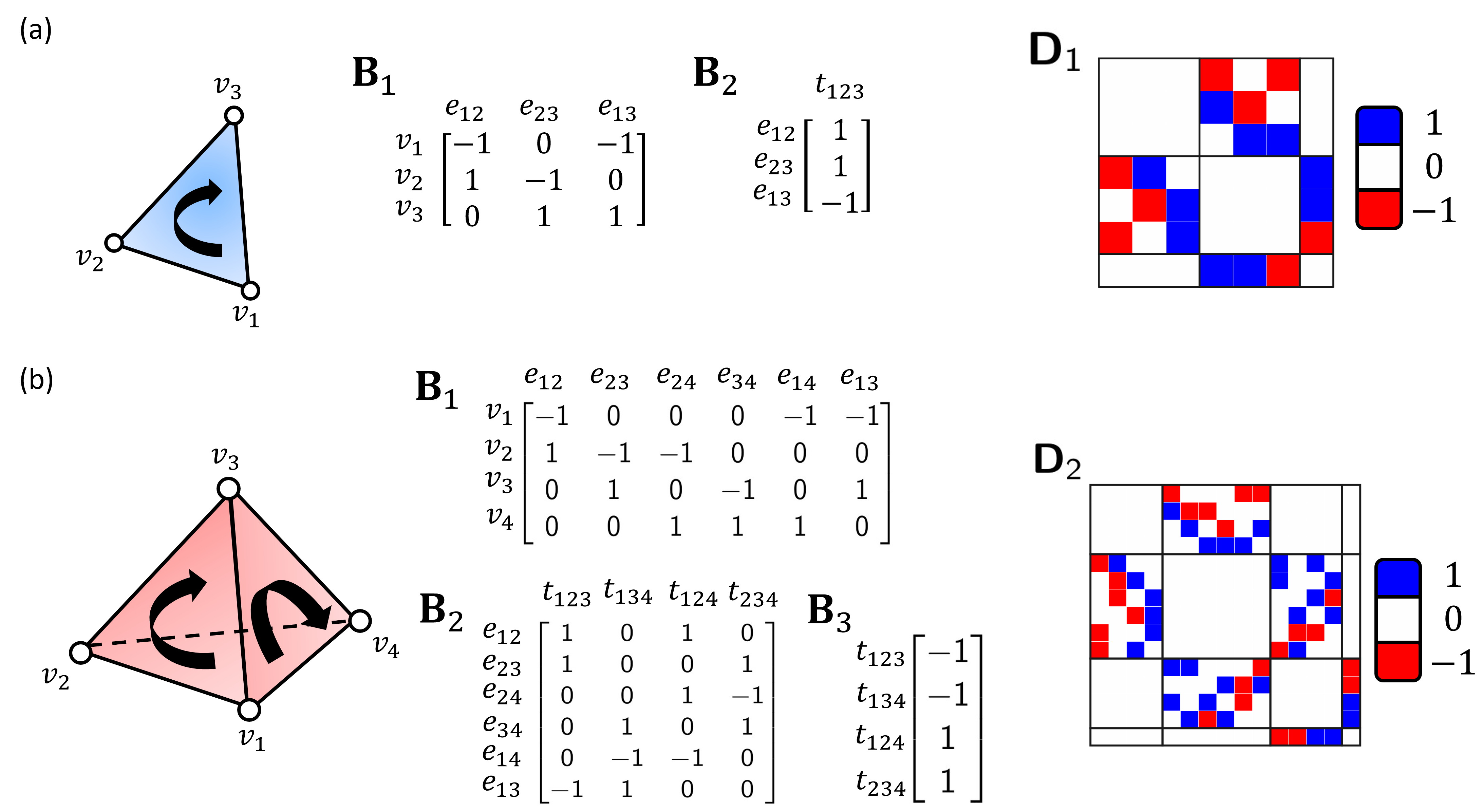}
	\caption{\label{fig:dirac_example}Illustration of constructions of (a): Discrete Dirac matrix $\mathbf{D}_1$ of a triangle and (b): Discrete Dirac matrix $\mathbf{D}_2$ of a tetrahedron along with its corresponding boundary matrices. The rows and columns of boundary matrices corresponds to a respective simplex each. For instance, in the boundary matrix $\mathbf{B}_2$ of Fig.~\ref{fig:dirac_example}(a), edge $e_{12}$ is oriented similarly as $t_{123}$, hence having an entry 1 in the matrix. As the entries of Dirac operator either take a value of $-1$, $0$ or $1$, the entries of Dirac operators are color coded with blue indicating $1$, white indicating $0$ and red indicating $-1$.}
\end{figure*}

\section*[Appendix]{Appendix B: Supplementary Details About Hodge Laplacian}
In this section, we show that (a): $\dim \ker \overline{\mathbf{L}}_p$ can be rewritten as $\dim \ker \overline{\mathbf{L}}_p^{\text{down}} - \dim \operatorname{im} \overline{\mathbf{L}}_p^{\text{up}}$ and in (b): Any eigenvector $v$ of $\overline{\mathbf{L}}_p$ can only either be $v \in \operatorname{im} \overline{\mathbf{L}}_p^{\text{up}} \subset \ker \overline{\mathbf{L}}_p^{\text{down}}$ or $v \in \operatorname{im} \overline{\mathbf{L}}_p^{\text{down}} \subset \ker \overline{\mathbf{L}}_p^{\text{up}}$.
\begin{proof}
	(a): Note that $\dim \ker \overline{\mathbf{L}}_p = \beta_p$ is similarly proven by Eckmann in 1944 \cite{eckmann1944harmonische}. Hence,
	\begin{align*}
	\dim \ker \overline{\mathbf{L}}_p = \beta_p &= \text{rank } Z_p - \text{rank } B_{p} \\
	&= \dim \ker \overline{\mathbf{B}}_p - \text{rank } \overline{\mathbf{B}}_{p+1}\\
	&= \dim \ker \overline{\mathbf{L}}_p^{\text{down}} - \text{rank } \overline{\mathbf{B}}_{p+1}^\top\\
	&= \dim \ker \overline{\mathbf{L}}_p^{\text{down}} - \dim C_p + \dim \ker \overline{\mathbf{B}}_{p+1}^\top \\
	&= \dim \ker \overline{\mathbf{L}}_p^{\text{down}} -  \dim C_p + \dim \ker \overline{\mathbf{L}}_p^{\text{up}}\\
	&= \dim \ker \overline{\mathbf{L}}_p^{\text{down}} - \dim \operatorname{im} \overline{\mathbf{L}}_p^{\text{up}}.	\end{align*}
	
	(b): $(\impliedby)$: For any non-zero eigenvalue $\lambda$ of $\overline{\mathbf{L}}_p^{\text{down}}$ (resp. $\overline{\mathbf{L}}_p^{\text{up}}$) with eigenvector $v$, (iv) and (v) (Appendix A) shows that $v \in \operatorname{im} \overline{\mathbf{L}}_p^{\text{up}}$ (resp. $v \in \im \overline{\mathbf{L}}_p^{\text{down}}$). Hence, $v \in \ker \overline{\mathbf{L}}_p^{\text{down}}$ or $v \in \ker \overline{\mathbf{L}}_p^{\text{up}}$. Then for both cases,
	\begin{align*}
	\overline{\mathbf{L}}_p v &= (\overline{\mathbf{L}}_p^{\text{down}} + \overline{\mathbf{L}}_p^{\text{up}})v =\overline{\mathbf{L}}_p^{\text{down}}v + \overline{\mathbf{L}}_p^{\text{up}}v= \lambda v.
	\end{align*}
	$(\implies)$: For any non-zero eigenvalue $\lambda$ of $\overline{\mathbf{L}}_p$,
	\begin{align*}
	\overline{\mathbf{L}}_pv &= \lambda v \implies \overline{\mathbf{L}}_p^{\text{down}}v + \overline{\mathbf{L}}_p^{\text{up}}v = \lambda v.
	\end{align*}
	From (v) and (vi) (Appendix A), a similar argument follows by showing that either $v \in \operatorname{im} \overline{\mathbf{L}}_p^{\text{up}} \subset \ker \overline{\mathbf{L}}_p^{\text{down}}$ or $v \in \operatorname{im} \overline{\mathbf{L}}_p^{\text{down}} \subset \ker \overline{\mathbf{L}}_p^{\text{up}}$.
\end{proof}

Additionally, let $\mathbf{s}(\overline{\mathbf{L}}_p^{\text{up}})$ and $\mathbf{s}(\overline{\mathbf{L}}_p^{\text{down}})$ be the spectrum of $\overline{\mathbf{L}}_p^{\text{up}}$ and $\overline{\mathbf{L}}_p^{\text{down}}$ respectively. Suppose the highest order of the simplicial complex $\mathcal{K}$ is $d$. Similar to \cite{horak2013spectra}, the multiplicity of zero eigenvalues in
\begin{itemize}
	\item[(i)] $\mathbf{s}(\overline{\mathbf{L}}_p^{\text{up}})$ can be computed as
	\begin{equation}\label{eqn:jost1}
	\dim C_p - \sum_{i=0}^{p} (-1)^{i+p} (\dim C_i - \dim H_i),
	\end{equation}
	\item[(ii)] $\mathbf{s}(\overline{\mathbf{L}}_p^{\text{down}})$ can be computed as
	\begin{equation}\label{eqn:jost2}
	\dim C_p - \sum_{i=0}^{p-1} (-1)^{p-1+i} (\dim C_i-\dim H_i).
	\end{equation}
\end{itemize}

\begin{proof}
(i): From Appendix A,
\begin{align*}
	\dim \ker \overline{\mathbf{L}}_p^{\text{up}} &= \dim \ker \overline{\mathbf{B}}_{p+1}^\top\\
	&= \dim C_{p} - \dim \text{im } \overline{\mathbf{B}}_{p+1}^\top\\
	&= \dim C_{p} - \dim \text{im } \overline{\mathbf{B}}_{p+1}.
\end{align*}

Then
\begin{align*}
	\dim \text{im } \overline{\mathbf{B}}_{p+1} &= \dim \ker \overline{\mathbf{B}}_{p} - \dim H_{p}\\
	&= \dim C_p - \dim H_p - \dim \text{im } \overline{\mathbf{B}}_{p}\\
	&= \cdots = \sum_{i=0}^{p} (-1)^{i+p} (\dim C_i - \dim H_i).
\end{align*}

Putting everything together yields
\begin{align*}
\ker \overline{\mathbf{L}}_p^{\text{up}}&=\dim C_{p} - \dim \text{im } \overline{\mathbf{B}}_{p+1} \\
&= \dim C_{p} - \sum_{i=0}^{p} (-1)^{i+p} (\dim C_i - \dim H_i).
\end{align*}

(ii): Since $\ker \overline{\mathbf{L}}_p^{\text{down}}=\ker \overline{\mathbf{B}}_p=\dim C_p - \dim \text{im }\overline{\mathbf{B}}_p$, then
\begin{align*}
	\dim \text{im }\overline{\mathbf{B}}_p &= \dim \ker \overline{\mathbf{B}}_{p-1} - \dim H_{p-1} \\
	&= \dim C_{p-1} - \dim H_{p-1} - \dim \text{im }\overline{\mathbf{B}}_{p-1}\\
	&= \cdots = \sum_{i=0}^{p-1} (-1)^{i+p-1} (\dim C_i - \dim H_i).
\end{align*}
Therefore,
\begin{align*}
	\ker \overline{\mathbf{L}}_p^{\text{down}}&=\dim C_p - \dim \text{im }\overline{\mathbf{B}}_p \\
	&= \dim C_p - \sum_{i=0}^{p-1} (-1)^{i+p-1} (\dim C_i - \dim H_i).
\end{align*}
\end{proof}

\section*[Appendix]{Appendix C: Proof for Hodge Decomposition}
\begin{proof}
 	Recall from Appendix A that $\ker(\overline{\mathbf{B}}_p)^\perp = \operatorname{im}(\overline{\mathbf{B}}_p^\top)$. Hence,
	\begin{align*}
	C_p &= \ker (\overline{\mathbf{B}}_p)\oplus \ker(\overline{\mathbf{B}}_p)^\perp \\
	&= \ker (\overline{\mathbf{B}}_p)\oplus \operatorname{im}(\overline{\mathbf{B}}_p^\top) \\
	&= \operatorname{im} (\overline{\mathbf{B}}_{p+1}) \oplus \ker(\overline{\mathbf{L}}_p) \oplus \operatorname{im}(\overline{\mathbf{B}}_p^\top),
	\end{align*}
	since $\ker(\overline{\mathbf{L}}_p) = H_p = \ker (\overline{\mathbf{B}}_p)/\operatorname{im} (\overline{\mathbf{B}}_{p+1})$.
\end{proof}

\section*[Appendix]{Appendix D: Proof that $\lambda^s$ is eigenvalue of Dirac operator}
\begin{proof}
	Since
	\begin{align}\label{eigen-1}
	\mathbf{D}_pv &= \lambda v,
	\end{align}
	then multiplying by $\mathbf{D}_p^{s-1}$ gives
	\begin{align*}
	\mathbf{D}_p^sv  &= \lambda \mathbf{D}_p^{s-1}v\\
	&= \lambda \mathbf{D}_p^{s-2}\mathbf{D}_pv\\
	&= \lambda^2 \mathbf{D}_p^{s-2}v = \cdots = \lambda^s v. \quad (\text{By } \eqref{eigen-1})
	\end{align*}
	Similarly, since $\mathbf{D}_pv = \lambda v$, we also have
	\begin{align}\label{eigen-2}
	\mathbf{D}_p\mathbf{Q}_pv &= -\lambda \mathbf{Q}_pv.
	\end{align}
	Hence,
	\begin{align*}
	\mathbf{D}_p^s\mathbf{Q}_pv  &= -\lambda \mathbf{D}_p^{s-1}\mathbf{Q}_pv\\
	&= -\lambda \mathbf{D}_p^{s-2}\mathbf{D}_p\mathbf{Q}_pv\\
	&= \lambda^2 \mathbf{D}_p^{s-2}\mathbf{Q}_pv = \cdots = (-\lambda)^s \mathbf{Q}_pv. \quad (\text{By } \eqref{eigen-2})
	\end{align*}
\end{proof}

\section*[Appendix]{Appendix E: Additional DO-based Fingerprints of OIHPs}
\begin{figure}[H]
	\centering
	\includegraphics[width=.4\textwidth]{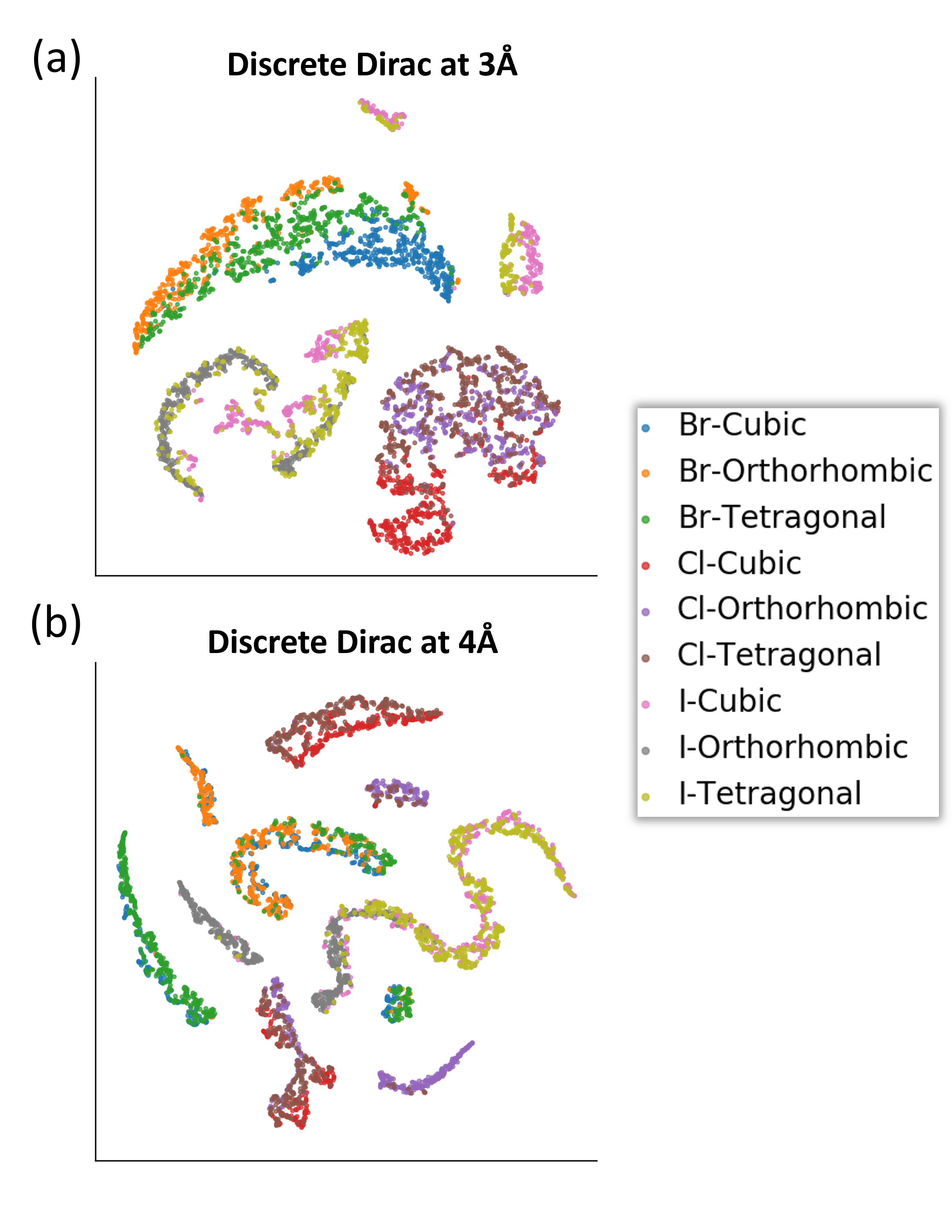}
	\caption{\label{fig:clustering_misc}The clustering of 9 types of OIHP molecular dynamics (MD) trajectories using DO-based features at other filtration times, namely, (A) Dirac operator at 3\AA\space and 4\AA. However, without persistence, the Dirac operators at specific cutoff distances only provide certain topological information. The x-axis and y-axis are the two principal components obtained from the $t$-SNE model.}
\end{figure}

\section*[Appendix]{Appendix F: Persistent Multiplicities of Hodge Laplacians}
\begin{figure}[H]
	\centering
	\includegraphics[width=.4\textwidth]{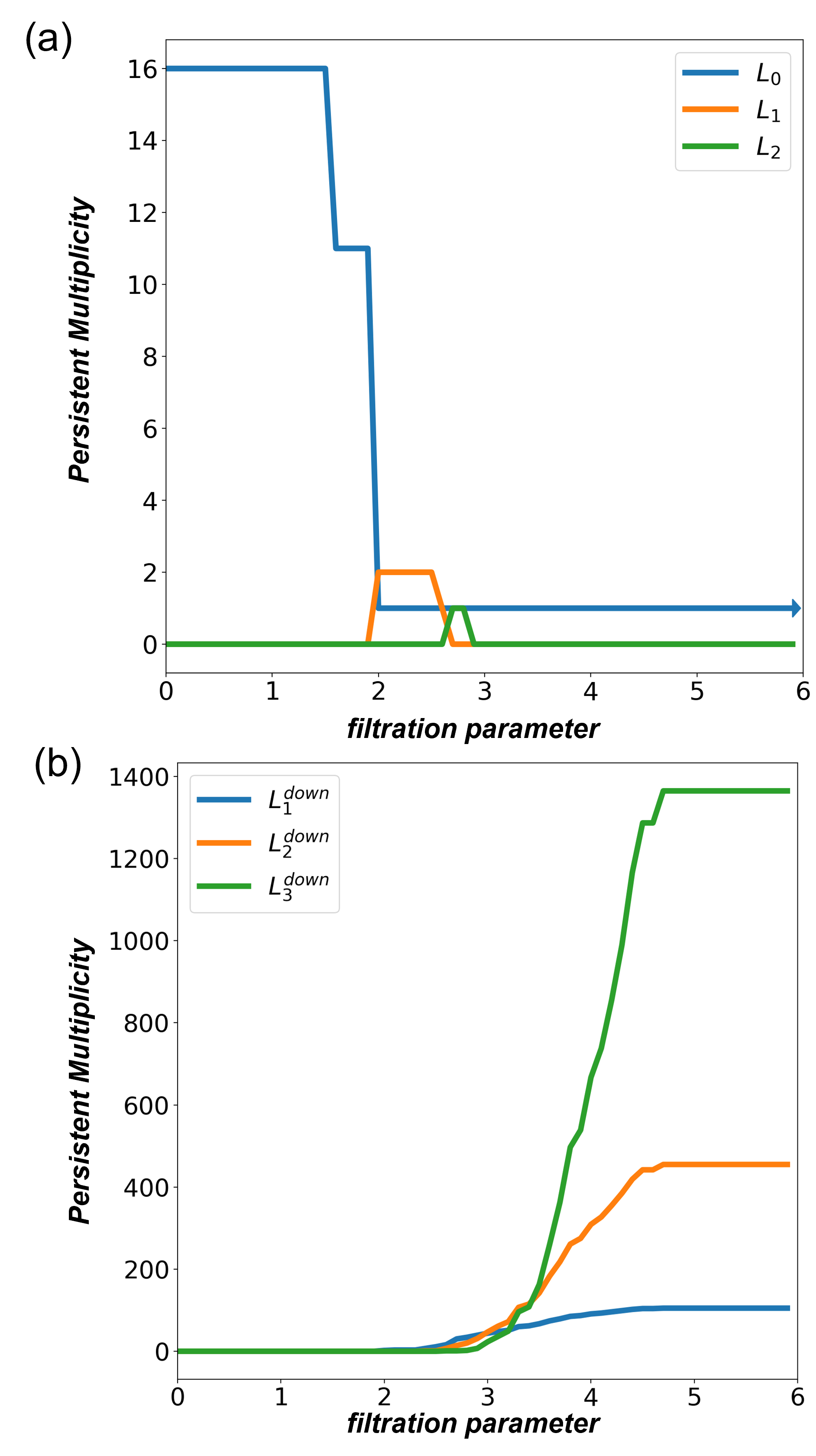}
	\caption{\label{fig:persmul_misc}The clustering of 9 types of OIHP molecular dynamics (MD) trajectories using DO-based features at other filtration times, namely, (A) Dirac operator at 3\AA\space and 4\AA. However, without persistence, the Dirac operators at specific cutoff distances only provide certain topological information. The x-axis and y-axis are the two principal components obtained from the $t$-SNE model.}
\end{figure}

\section*[Appendix]{Appendix G: Spectrum of Dirac Matrices}
Let $\mathbf{A}_1$ be an $m_1 \times n_1$ matrix and $\mathbf{A}_2$ to be a $m_2 \times n_2$ matrix,
\[
\textbf{diag}(\mathbf{A}_1, \mathbf{A}_2) = \begin{bmatrix}
\mathbf{A}_1 & \bzero_{m_1\times n_2} \\
\bzero_{m_2\times n_1}  &\mathbf{A}_2
\end{bmatrix}.
\]
This gives
\begin{align*}
\mathbf{D}_p^2 &= \textbf{diag}(\mathbf{L}_0,\mathbf{L}_1, \cdots, \mathbf{L}_{p+1}) \\
&= \textbf{diag}(\mathbf{L}_0^{\text{up}}, \mathbf{L}_1^{\text{up}},\cdots, \mathbf{L}_{p+1}^{\text{up}})\\
& \quad+ \textbf{diag}(\mathbf{L}_0^{\text{down}}, \mathbf{L}_1^{\text{down}},\cdots, \mathbf{L}_{p+1}^{\text{down}}),
\end{align*}
where $\mathbf{L}_{p+1}^{\text{up}} = \bzero_{n_{p+1}\times n_{p+1}}$ and $\mathbf{L}_0^{\text{down}} = \bzero_{n_{0}\times n_0}$. For convenience, we simply denote $\textbf{diag}(\mathbf{L}_0^{\text{up}}, \mathbf{L}_1^{\text{up}},\cdots, \mathbf{L}_{p+1}^{\text{up}})$ as $(\mathbf{D}_p^2)^{\text{up}}$ and $\textbf{diag}(\mathbf{L}_0^{\text{down}}, \mathbf{L}_1^{\text{down}},\cdots, \mathbf{L}_{p+1}^{\text{down}})$ as $(\mathbf{D}_p^2)^{\text{down}}$.

Let $\mathbf{s}(\mathbf{D}_p^2)^{\text{up}}$ and $\mathbf{s}(\mathbf{D}_p^2)^{\text{down}}$ be the spectrum of the upper and lower $\mathbf{D}_p^2$. Following \eqref{eqn:jost1} and \eqref{eqn:jost2}, the multiplicity of zero eigenvalues for
\begin{itemize}
	\item[(i)] $\mathbf{s}(\mathbf{D}_p^2)^{\text{up}}$ can be computed as
	\begin{equation*}
	\sum_{k=0}^{p+1}\bigg(\dim C_k - \sum_{i=0}^{k} (-1)^{i+k} (\dim C_i - \dim H_i)\bigg).
	\end{equation*}
	\item[(ii)] $\mathbf{s}(\mathbf{D}_p^2)^{\text{down}}$ can be computed as
	\begin{equation*}
	\sum_{k=0}^{p+1} \bigg(\dim C_k - \sum_{i=0}^{k-1} (-1)^{k-1+i} (\dim C_i-\dim H_i).\bigg).
	\end{equation*}
\end{itemize}
Furthermore, we also have
\begin{align*}
\dim \ker \mathbf{D}_p^2 &= \sum_{k=0}^{p+1} \beta_k = \sum_{k=0}^{p+1} (\dim\ker \mathbf{L}_k^{\text{down}}- \dim\operatorname{im}\mathbf{L}_k^{\text{up}})\\
&= \sum_{k=0}^{p+1} \dim\ker \mathbf{L}_k^{\text{down}} - \sum_{k=0}^{p+1} \dim\operatorname{im}\mathbf{L}_k^{\text{up}}\\
&= \dim \ker (\mathbf{D}_p^2)^{\text{up}} - \dim \operatorname{im} (\mathbf{D}_p^2)^{\text{down}}.
\end{align*}

\section*[Appendix]{Derivation of the real spectrum of the Weighted Dirac operator}

By setting $\mathbf{G}_n$ as identity matrices, we obtain the Dirac matrices whose eigenvalues has been shown to be always real \cite{calmon2023dirac}. In \cite{calmon2023dirac}, the special case is discussed and has been applied to signal processing by proposing the use of topological spinors obtained from eigenspectrum of Dirac matrices. Essentially, an $n$-dimensional topological spinor $\mathbf{s}$ can be written as
\[
\mathbf{s} = \begin{bmatrix}
\mathbf{s}_0 \\
\mathbf{s}_1 \\
\vdots\\
\mathbf{s}_n \\
\end{bmatrix} \in \mathcal{C}_n,
\] 
where $\mathcal{C}_n$ is the space of all $n$-dimensional topological spinors. Here, the $n$-dimensional topological spinor $\mathbf{s}$ is a direct sum of block vectors (signals) $\mathbf{s}_k$ defined for $k$-simplices, $0\leq k \leq n$. 
Now, we shall provide a similar treatment to the weighted Dirac matrices which may or may not be symmetric but can be shown to always have real eigenvalues.  To be concrete we will focus on the case in which the simplicial complex is two dimensional, i.e. formed by nodes, links and triangles. Extension of these results to higher-order Dirac operators is straightforward.

Recall from the definition of weighted Dirac matrix that $\overline{\mathbf{D}}_1$ is written as 
\begin{equation*}
\label{eqn:metricdirac2}
\renewcommand{\arraystretch}{1.5}
\overline{\mathbf{D}}_1=\begin{bmatrix}
\bzero_{n_0\times n_0}    & \mathbf{G}_{0}^{-1}\mathbf{B}_1\mathbf{G}_{1}/\sqrt{2}    & \bzero_{n_0\times n_2}     \\

\mathbf{B}_1^\top/\sqrt{2}  & \bzero_{n_1\times n_1}      & \mathbf{G}_{1}^{-1}\mathbf{B}_2\mathbf{G}_{2}/\sqrt{3}\\ 

\bzero_{n_2\times n_0}      & \mathbf{B}_2^\top/\sqrt{3}  & \bzero_{n_2\times n_2}\\
\end{bmatrix}
\end{equation*}

We shall write $\overline{\mathbf{D}}_1$ as $\overline{\mathbf{D}}_{[0]}+\overline{\mathbf{D}}_{[1]}$ where 
\begin{equation*}
\label{eqn:metricdirac3}
\renewcommand{\arraystretch}{1.5}
\overline{\mathbf{D}}_{[0]}=
\begin{bmatrix}
\bzero_{n_0\times n_0}    & \mathbf{G}_{0}^{-1}\mathbf{B}_1\mathbf{G}_{1}/\sqrt{2}    & \bzero_{n_0\times n_2}     \\

\mathbf{B}_1^\top/\sqrt{2}  & \bzero_{n_1\times n_1}      & \bzero_{n_1\times n_2} \\ 

\bzero_{n_2\times n_0}      & \bzero_{n_2\times n_1}  & \bzero_{n_2\times n_2}\\
\end{bmatrix}
\end{equation*}
and
\begin{equation*}
\label{eqn:metricdirac4}
\renewcommand{\arraystretch}{1.5}
\overline{\mathbf{D}}_{[1]}=
\begin{bmatrix}
\bzero_{n_0\times n_0}    & \bzero_{n_0\times n_1}    & \bzero_{n_0\times n_2}     \\

\bzero_{n_1\times n_0}   & \bzero_{n_1\times n_1}      & \mathbf{G}_{1}^{-1}\mathbf{B}_2\mathbf{G}_{2}/\sqrt{3}\\ 

\bzero_{n_2\times n_0}      & \mathbf{B}_2^\top/\sqrt{3}  & \bzero_{n_2\times n_2}\\
\end{bmatrix}
\end{equation*}

Note that $\overline{\mathbf{D}}_{[1]}\overline{\mathbf{D}}_{[0]}=\bzero$ and $\overline{\mathbf{D}}_{[0]}\overline{\mathbf{D}}_{[1]}=\bzero$, 
which implies that 
\[
\text{im } \overline{\mathbf{D}}_{[1]} \subseteq \ker \overline{\mathbf{D}}_{[0]}, \quad \text{im } \overline{\mathbf{D}}_{[0]} \subseteq \ker \overline{\mathbf{D}}_{[1]}.
\]
This means that the weighted Dirac matrix $\overline{\mathbf{D}}_1$ admits the following Dirac decomposition \cite{calmon2023dirac}:
\[
\mathcal{C}_2 = \ker \overline{\mathbf{D}}_1 \oplus \text{im } \overline{\mathbf{D}}_{[0]} \oplus \text{im }\overline{\mathbf{D}}_{[1]},
\]
where
\[
\ker \overline{\mathbf{D}}_1 = \ker \overline{\mathbf{L}}_{[0]}\oplus \ker \overline{\mathbf{L}}_{[1]}\oplus\ker \overline{\mathbf{L}}_{[2]}.
\]

The above Dirac decomposition implies that the non-zero eigenvectors of $\overline{\mathbf{D}}_1$ are either non-zero eigenvectors of $\overline{\mathbf{D}}_{[0]}$ (corresponding to an eigenvalue $\lambda_0$) or non zero eigenvectors of $\overline{\mathbf{D}}_{[1]}$ (corresponding to an eigenvalue $\lambda_1$).
Here, define the matrix $\Phi$ of the eigenvectors of $\overline{\mathbf{D}}_1$ as 
\bea
\bm \Phi = 
\begin{bmatrix}\bm \Phi_0& \bm \Phi_1 &\bm \Phi_{\text{harm}}\nonumber \\
\end{bmatrix}.
\eea
where ${\bm \Phi}_n$ is the matrix of the eigenvectors ${\bm\phi}_n\in \mbox{im}(\overline{\mathbf{D}}_{[n]})$ with $n\in \{0,1\}$ and $\bm\Phi_{\text{harm}}$ is the matrix of eigenvectors forming a basis for $\mbox{ker}(\overline{\mathbf{D}}_1)$. 

Now, denote ${\bf u}_0$ as the eigenvector of ${\bf L}_{[0]}$ and  ${\bf v}_0$ as the eigenvector of ${\bf L}_{[1]}^{\text{down}}$ corresponding to the same non zero eigenvalue $\Lambda_0$, i.e. satisfying the relations
\bea
\mathbf{L}_{[0]}{\bf u}_0=\Lambda_0{\bf u}_0,\quad
\mathbf{L}_{[1]}^{\text{down}}{\bf v}_0=\Lambda_0{\bf v}_0\nonumber
\eea
and similarly, we have ${\bf v}_1$ as the eigenvector of ${\bf L}_{[1]}^{\text{up}}$ and ${\bf z}_1$ as the eigenvector of ${\bf L}_{[2]}^{\text{down}}$ corresponding to a same non zero eigenvalue $\Lambda_1$, i.e. satisfying the relations
\bea
\mathbf{L}_{[1]}^{\text{up}}{\bf v}_1=\Lambda_1{\bf v}_1,\quad
\mathbf{L}_{[2]}^{\text{down}}{\bf z}_1=\Lambda_1{\bf z}_1,\nonumber
\eea
with ${\bf u}_0,{\bf v}_0,{\bf v}_1$ and ${\bf z}_1$ being eigenvectors normalized to one.
Using a notation from Appendix G, we can also write $\overline{\mathbf{D}}_{p}^2 = \textbf{diag}(\mathbf{L}_{[0]},\mathbf{L}_{[1]}, \cdots, \mathbf{L}_{[p+1]})$. This implies that
eigenvectors $\overline{\mathbf{D}}_{[0]}$ and the eigenvectors of $\overline{\mathbf{D}}_{[1]}$ takes the form 
\bea
&\bm \Phi_0 = 
\begin{bmatrix}{\bf U}_{0} &  {\bf U}_{0}  \\
	{\bf V}_0  & -{\bf V}_0 \\
	\bm 0&\bm 0
\end{bmatrix},
\bm \Phi_1 = 
\begin{bmatrix}\bm 0&\bm 0\\{\bf V}_1 &  {\bf V}_1  \\
	{\bf Z}_1  & -{\bf Z}_1 \\
\end{bmatrix}\nonumber
\eea
where ${\bf U}_0,{\bf V}_0,{\bf V}_1,{\bf Z}_1$ are the matrices formed by vectors proportional to the eigenvectors ${\bf u}_0,{\bf v}_0,{\bf v}_1$ and ${\bf z}_1$ respectively.
In particular we have that the eigenvectors $\bm {\phi_n}$ with $n\in \{0,1\}$ can be written as 
\bea
\bm \phi_0^+=\frac{1}{\mathcal{N}_0}\left(\begin{array}{cc} {\bf u}_0\\  {\bf v}_0&\\{\bf 0}\end{array}\right),\quad \bm \phi_0^-=\frac{1}{\mathcal{N}_0}\left(\begin{array}{cc} {\bf u}_0\\  -{\bf v}_0&\\{\bf 0}\end{array}\right)
\nonumber \\
\bm \phi_1^+=\frac{1}{\mathcal{N}_1}\left(\begin{array}{cc}{\bf 0}\\{\bf v}_1 \\   {\bf z}_1 \end{array}\right), \quad \bm \phi_1^-=\frac{1}{\mathcal{N}_1}\left(\begin{array}{cc}{\bf 0}\\{\bf v}_1 \\   -{\bf z}_1 \end{array}\right). \nonumber 
\eea

Let us indicate with ${\bf u}_{\text{harm}},{\bf v}_{\text{harm}}$ and ${\bm z}_{\text{harm}}$ the eigenvectors corresponding to the zero eigenvalue of ${\bf L}_{[0]},{\bf L}_{[1]}$ and ${\bf L}_{[2]}$ respectively, i.e. satisfying
\bea
{\bf L}_{[0]}{\bf u}_\text{harm}&=&{\bf 0},\nonumber \\
{\bf L}_{[1]}{\bf v}_\text{harm}&=&({\bf L}_{[1]}^{\text{up}}+{\bf L}_{[1]}^{\text{down}}){\bf v}_\text{harm}={\bf 0},\nonumber \\
{\bf L}_{[2]}{\bf z}_\text{harm}&=&{\bf 0}.\nonumber
\eea
We have that 
\bea
\bm \Phi_\text{harm} = 
\begin{bmatrix} {\bm U}_{\text{harm}} & \bm 0 &\bm 0  \\
	\bm 0& {\bm V}_\text{harm}  & \bm 0 \\
	\bm 0&\bm 0 & {\bm Z}_\text{harm}\end{bmatrix},\nonumber
\eea 
where ${\bm U}_{\text{harm}},{\bm V}_{\text{harm}}$ and ${\bm Z}_{\text{harm}}$ are the matrices of eigenvectors ${\bf u}_{\text{harm}},{\bf v}_{\text{harm}}$ and ${\bf z}_{\text{harm}}$  respectively.
The weighted Dirac matrix has eigenvalues which can be null, positive or negative. The positive part of the spectrum is given by the square root of the eigenvectors of the (normalized) Hodge Laplacian and for each positive eigenvector there is a negative eigenvector with the same absolute value.
The eigenvectors of the weighted Dirac matrix are formed by the direct sum of the eigenvectors of the weighted Hodge Laplacians. 

The normalized weighted Dirac matrix has positive, zero and negative eigenvalues $\lambda_n$ that have absolute value smaller or equal to one \cite{calmon2023dirac}
\begin{equation}
|\lambda_n|\leq 1.\nonumber
\end{equation}
with $\lambda_n$ related to the eigenvalues $\Lambda_{n-1}$ by
\bea
\lambda_n=\pm \sqrt{\Lambda_{n-1}}.
\eea
Now let us define 
\bea
\tilde{\bf B}_n={\bf G}_{n-1}^{-1/2}{{\bf B}}_n{\bf G}_{n}^{1/2}/\sqrt{n+1}\nonumber
\eea
and equivalently its transpose
\bea
\tilde{\bf B}_n^{\top}={\bf G}_{n}^{1/2}{{\bf B}}_n^{\top}{\bf G}_{n-1}^{-1/2}/\sqrt{n+1}.\nonumber
\eea
whose product leads to the symmetric normalized Hodge Laplacians $\tilde{\bf L}_{[n-1]}^{\text{up}}=\tilde{\bf B}_n\tilde{\bf B}_n^{\top}$ and $\tilde{\bf L}_{[n]}^{\text{down}}=\tilde{\bf B}_n^{\top}\tilde{\bf B}_n$. 
From this definition it follows that 
\bea
\tilde{\bf L}_{[n-1]}^{\text{up}}&=&{\bf G}_{n-1}^{1/2}{\bf L}_{[n]}^{\text{up}}{\bf G}_{n-1}^{-1/2}\nonumber \\
\tilde{\bf L}_{[n]}^{\text{down}}&=&{\bf G}_{n}^{1/2}{\bf L}_{[n]}^{\text{down}}{\bf G}_{n}^{-1/2} 
\eea
By defining $\tilde{\Lambda}_n$ as the eigenvalues satisfying 
\bea
\tilde{\mathbf{L}}_{[0]}\tilde{\bf u}_0=\tilde\Lambda_0\tilde{\bf u}_0,\quad
\tilde{\mathbf{L}}_{[1]}^{\text{down}}\tilde{\bf v}_0=\tilde\Lambda_0\tilde{\bf v}_0\nonumber \\
\tilde{\mathbf{L}}_{[1]}^{\text{up}}\tilde{\bf v}_1=\tilde\Lambda_1\tilde{\bf v}_2,\quad
\tilde{\mathbf{L}}_{[2]}^{\text{down}}\tilde{\bf z}_1=\Lambda_1\tilde{\bf z}_1,
\eea
It is easy to show that the 
eigenvalues $\Lambda_n$ are  equal to the eigenvalues $\tilde{\Lambda}_n$, i.e.
\bea
\Lambda_n=\tilde\Lambda_n=|\mu_n|^2\nonumber
\eea
where $\mu_n$ is the singular value of 
$\tilde{\bf B}_n$ and that 
\bea
{\bf u}_0={\bf G}_0^{-1/2}\tilde{\bf u}_0\quad {\bf v}_0={\bf G}_1^{-1/2}\tilde{\bf v}_0\nonumber \\
{\bf v}_1={\bf G}_1^{-1/2}\tilde{\bf v}_1\quad {\bf z}_1={\bf G}_2^{-1/2}\tilde{\bf z}_1.
\eea
It follows that the spectrum of the normalized Dirac operator is real although the operator is not symmetric with 
\bea
\lambda_n=\pm |\mu_n|
\eea
and the eigenvectors $\bm\phi^{\pm}_0$ and $\bm\phi^{\pm}_1$ are given by 
\bea
\bm \phi_0^+=\frac{1}{\mathcal{N}_0}\left(\begin{array}{cc} {\bf G}_0^{-1/2}\tilde{\bf u}_0\\  {\bf G}_1^{-1/2}\tilde{\bf v}_0&\\{\bf 0}\end{array}\right),\quad \bm \phi_0^-=\frac{1}{\mathcal{N}_0}\left(\begin{array}{cc} {\bf G}_0^{-1/2}\tilde{\bf u}_0\\  -{\bf G}_1^{-1/2}\tilde{\bf v}_0&\\{\bf 0}\end{array}\right)
\nonumber \\
\bm \phi_1^+=\frac{1}{\mathcal{N}_1}\left(\begin{array}{cc}{\bf 0}\\{\bf G}_1^{-1/2}\tilde{\bf v}_1 \\   {\bf G}_2^{-1/2}\tilde{\bf z}_1 \end{array}\right), \quad \bm \phi_1^-=\frac{1}{\mathcal{N}_1}\left(\begin{array}{cc}{\bf 0}\\{\bf G}_1^{-1/2}\tilde{\bf v}_1 \\   -{\bf G}_2^{-1/2}\tilde{\bf z}_1 \end{array}\right), \nonumber 
\eea
where $\tilde{\bf u}_0, \tilde{\bf v}_0$ are the left and the right singular vectors of $\tilde{\mathbf{B}}_1$ respectively and where $\tilde{\bf v}_1, \tilde{\bf z}_1$ are the left and the right singular vectors of $\tilde{\mathbf{B}}_2$ respectively.
\end{document}